\documentclass[11pt]{article}
\usepackage{epsf}
\def\ltap{\ \raisebox{-.4ex}{\rlap{$\sim$}} \raisebox{.4ex}{$<$}\ }
\def\gtap{\ \raisebox{-.4ex}{\rlap{$\sim$}} \raisebox{.4ex}{$>$}\ }

\begin{document}


\begin{titlepage}

\begin{flushright}
RESCEU-UT-3/01      \\
SISSA 26/01/EP      \\ 
IISc-CTS/12/01      \\ 
PM/01-22            \\
TU-621              \\
hep-ph/0108244      \\ 
\end{flushright}
\vspace{0.5cm}
\begin{center}
{\Large
{\bf  Lightest-neutralino decays in $R_p$-violating models with}     \\[1.01ex]
{\bf  dominant $\lambda^{\prime}$ and $\lambda$ couplings}}          \\[5ex]
 F.~Borzumati$^{\,a,b}$, R.M.~Godbole$^{\,c}$, 
  J.L.~Kneur$^{\,d}$, and F.~Takayama$^{\,e}$                        \\[1ex]
{\it ${}^a$   RESCEU, University of Tokyo, Tokyo 113-0033, Japan}         
{\it ${}^b$   SISSA, Via Beirut 4, I--34014 Trieste, Italy} 
{\it ${}^{c}$ CTS, Indian Institute of Science, Bangalore 560-012, India}
{\it ${}^{d}$ LPMT, Universit\'e de Montpellier II,
                  F--34095 Montpellier Cedex 5, France}
{\it ${}^{e}$ Department of Physics, Tohoku University, Sendai 980-8578, Japan}  
\end{center}

\vspace{1.5cm}
\begin{center} 
ABSTRACT \\
\vspace*{8mm}
\parbox{15cm}{Decays of the lightest neutralino are studied in
$R_p$-violating models with operators $\lambda^\prime L Q D^c$ and
$\lambda L L E^c$ involving third-generation matter fields and with
dominant $\lambda^\prime$ and $\lambda$ couplings.  Generalizations to
decays of the lightest neutralino induced by subdominant
$\lambda^\prime$ and $\lambda$ couplings are straightforward.  Decays
with the top-quark among the particles produced are considered, in
addition to those with an almost massless final state.
Phenomenological analyses for examples of both classes of decays are
presented.  No specific assumption on the composition of the lightest
neutralino is made, and the formulae listed here can be easily
generalized to study decays of heavier neutralinos.  It has been recently
pointed out that, for a sizable coupling $\lambda^\prime_{333}$,
tau-sleptons may be copiously produced at the LHC as single
supersymmetric particles, in association with top- and bottom-quark
pairs. This analysis of neutralino decays is, therefore, a first step
towards the reconstruction of the complete final state produced in this
case.}
\end{center}

\vfill
\end{titlepage}
\thispagestyle{empty}

%
\section{Introduction} 
\label{intro}

Within the Standard Model (SM), electroweak gauge invariance is
sufficient to ensure conservation of both the lepton number $L$ and
the baryon number $B$, at least in a perturbative context. This is not
the case in supersymmetric theories, wherein it is possible to write
terms in the superpotential that are invariant under supersymmetric
and SM gauge transformations, but that do not conserve $B$, nor $L$.
These terms violate a discrete
symmetry~\cite{Fayet:1977yc,Farrar:1978xj}, called $R$-parity, $R_p$,
which implies a conserved quantum number $R_p \equiv (-1)^{3B+L+2S}$,
where $S$ stands for the spin of the particle.  It is clear from this
definition that all the SM particles have $R_p = +1$, whereas all
superpartners have $R_p = -1$. It is this discrete symmetry that
guarantees the stability of the lightest supersymmetric particle (LSP)
and thus also supplies a candidate for the cold dark matter. The
$R_p$-violating terms in the superpotential can be written as
\begin{equation}
W_{{\not{R}}_p} =\frac{1}{2} \lambda''_{ijk} U^c_i D^c_j D^c_k -
\frac{1}{2} \lambda_{ijk} L_i L_j E^c_k - \lambda'_{ijk} L_i Q_j D^c_k
+ \kappa_i L_i H_2 \,,
\label{superpot}
\end{equation}
where $L_i,Q_i$ represent the SU(2) lepton and quark doublet
superfields, $E^c_i,D^c_i$ and $U^c_i$, the SU(2) lepton and quark
singlet superfields, and $i,j,k$ are generation indices.  The symbol
$H_2$ denotes the Higgs superfield that gives rise to the entries in
the up-type quark mass matrix, through the Yukawa term $h_U H_2 Q U^c$. 

Imposition of $R_p$ invariance gets rid of the potentially dangerous
terms in eq.~(\ref{superpot}) that can cause a too fast proton decay.
To this aim, however, it is sufficient to forbid the simultaneous
presence of the $B$-violating $\lambda^{\prime \prime}$ terms and the
$L$-violating $\lambda^{\prime}$ terms. The existence of $B$-violating
terms that are not negligibly small can wash out
baryogenesis~\cite{erase,Ma:1999ni}.  (Baryogenesis may be triggered
by leptogenesis through nonperturbative effects. In this case,
leptogenesis requires that at least one combination of the
$L$-violating couplings is very
small~\cite{Dreiner:1993vm,Ma:1999ni}.)  Furthermore, in studies of
unified string theories it was shown~\cite{Ibanez:1992pr} that there
exist discrete symmetries which treat leptons and quarks differently,
as $R_p$ does. In particular, in this context, given the particle
content of the minimal supersymmetric Standard Model
(MSSM)~\cite{MSSM}, the lack of rapid proton decay implies two
symmetries: $R_p$ and $B$. Imposing the first eliminates only the
dimension-\emph{four} operators that cause proton decay, whereas $B$
conservation removes also the dimension-\emph{five} operators. Thus,
the option of $B$ conserved and $R_p$ broken through $L$ violation
seems the theoretically interesting one, and it is the one adopted in
this paper.

Data from neutrino oscillation experiments show now clear evidence for
nonvanishing neutrino masses~\cite{Groom:2000in}. $R_p$-violating
models, with their new interaction terms, provide an alternative way
to generate these masses at the tree and quantum
level~\cite{RPV-NUM1}--\cite{RPV-NUM3} consistent with all
informations from the different oscillation experiments, without
having to introduce new superfields in addition to those of the
MSSM~\cite{RPV-NUEXP}.

This discussion makes it clear that studies of $R_p$-violating models
are interesting from a theoretical point of view. A much more
pragmatic interest in them is due to the instability of the LSP
induced by $R_p$-violating couplings.  If at least one of the
$R_p$-violating couplings is $> 10^{-6}$ or so, then, the LSP decays
within the detector~\cite{Dawson:1985vr}.  The instability of the LSP
changes the phenomenology of sparticle searches at colliders quite
drastically (see, among others, ref.~\cite{Godbole:1993fb}), as it
changes the type, the number and the energies of the final state
particles. If not too small, these couplings may give rise to
interesting collider signatures, other than those due to the LSP decay
(see for example ref.~\cite{Borzumati:1999th}).

The flip side of considering $R_p$-violating models, of course, is the
existence of additional unknown parameters, as many as 48 in the
superpotential. Unfortunately, there is no theoretical indications
about their size, as for their $R_p$-conserving counterparts, the
Yukawa couplings. Lacking any theoretical guideline about the
magnitude of these couplings, one can rely only on phenomenological
constraints obtained in various experiments to get a clue to their
size.  As discussed in section~\ref{limits}, the dimensionful
couplings $\kappa_i$ are severely restricted by tree-level
contributions to neutrino masses. Dimensionless trilinear couplings
are, in general, also constrained. If one excludes neutrino data,
however, the restrictions are less severe for couplings with at least
one third-generation index.  It is, therefore, not unreasonable to
expect that the former are the largest couplings, as the corresponding
third generation Yukawa couplings are larger than the first and second
generation ones.

Hence, it is an interesting question to ask what role can LHC play in
probing these couplings. In ref.~\cite{Borzumati:1999th}, it was
pointed out that a sizable $\lambda^\prime_{333}$ can give rise to the
process $p p \to t \tilde{\tau} X$, or $p p \to t \bar{b}
\tilde{\tau}X$, depending on how many $b$-quarks in the final state
can actually be tagged. (In the following, for the sake of clarity, we
shall neglect this complication and list final states with all
produced $b$-quarks.) The production cross section is, for example,
$\sim 10 fb$ for $m_{\tilde{\tau}}$ up to $1\,$TeV, if
$\lambda^\prime_{333} \sim 0.5$.  Such a large value of
$\lambda^\prime_{333}$ may control also the decay of $\tilde{\tau}$-slepton,
as well as that of the lightest neutralino and chargino, thus
providing very distinctive final states for the signal.  To be
specific, the slepton $\tilde{\tau}$ can decay into the lightest
neutralino, $\tilde{\chi}^0_1$, and, if kinematically allowed, in the
lightest chargino, $\tilde{\chi}^-_1$, through gauge interactions:
\begin{equation}
\tilde{\tau}  \to  \tau \tilde{\chi}^0_1\,, \qquad 
\tilde{\tau}  \to  \nu_\tau \tilde{\chi}^-_1\,.
\label{staudecGI}
\end{equation}
If heavy enough, it may also decay with a substantial rate into:
\begin{equation}
\tilde{\tau} \to b \bar{t}\,,
\label{staudecRP}
\end{equation}
through the $R_p$-violating interaction $\lambda'_{333} L_3 Q_3
D^c_3$.\footnote{If the slepton $\tilde{\tau}$ is the LSP, and light,
as advocated for example in~\cite{ALS}, $\tilde{\chi}^0_1$,
$\tilde{\chi}^-_1$, and possibly also $t$, are off-shell in
eqs.~(\ref{staudecGI}) and~(\ref{staudecRP}).}  In turn, the lightest
neutralino obtained from $\tilde{\tau}$, as in eq.~(\ref{staudecGI}),
decays as:
\begin{equation}
\tilde{\chi}^0_1  \to  b \bar{b}\nu_\tau\,, \qquad 
\tilde{\chi}^0_1  \to  t \bar{b}\tau\,,  \qquad 
\tilde{\chi}^0_1  \to  \bar{t} b\bar{\tau}\,, 
\label{neudecRP}
\end{equation}
with the second and third decay mode allowed only for a heavy
$\tilde{\chi}^0_1$, i.e.\ for $m_{\tilde{\chi}^0_1} > m_t + m_b
+m_\tau$. One should notice that in the case of the decay of
eq.~(\ref{staudecRP}), the final signature
\begin{equation}
pp \to t \bar{b} \,\tilde{\tau} X \to t \bar{b} \,\bar{t} b X\,,
\label{Higgsignal} 
\end{equation}
is also that obtained from a charged Higgs boson\footnote{In
$R_p$-violating models, the $\tilde{\tau}$-sleptons are not
necessarily lighter than the charged Higgs bosons, and they may
therefore give similar contributions to the above cross section, when
their couplings to third-generation quarks are of the same order of
magnitude of the $b$- and $t$-quark Yukawa couplings. Moreover, their
masses are not constrained by low-energy processes such as $b \to s
\gamma$~\cite{RPBSGAMMA}, in sharp contrast to the situation for the
charged-Higgs boson mass in $R_p$-conserving
models~\cite{cHBSGAMMA}. See also discussion in the next section.}
produced in association with a $t$- and $b$-quark pair through
$R_p$-conserving interactions~\cite{Borzumati:1999th,chHiggs}, or from
pair-produced charged Higgs bosons. On the contrary, the final states
obtained when the $\tilde{\tau}$-slepton decays, for example, as $\tilde{\tau} \to
\tau \tilde{\chi}^0_1$, i.e.,
\begin{equation}
pp \to t \bar{b} \,\tilde{\tau} X \to (2t)\,(2\bar{b}) \,(2\tau) X \,,
\qquad 
t \bar{t} \,b \bar{b} \,\tau \bar{\tau} X\,, 
\qquad {\rm and} \quad 
t b \,(2\bar{b}) \,\tau \nu_\tau X\,,
\label{staurpsignals} 
\end{equation}
cannot be confused with those originated by a charged Higgs boson, if
more than $2b$'s can be tagged.  In particular, the first final state
in eq.~(\ref{staurpsignals}) is quite distinctive of $R_p$-violating
models since it gives rise to a lepton-number violating final state,
with two like-sign $\tau$'s. This is an identifying signal of
singly-produced $\tau$-sleptons, when the $W^+$ bosons obtained from
the $t$-quarks decay hadronically.

Two like-sign $\tau$'s will be also produced by heavy lightest
neutralinos pair-produced at an $e^+ e^-$
collider,\footnote{Like-sign dileptons were recognized since long as
the typical signal for the decay of a pair of lightest neutralinos
into light fermions in $R_p$-violating models. See for example
ref.~\cite{Godbole:1993fb}.} when $\lambda^\prime_{333}$ is much
larger than all other $R_p$-violating couplings. Under this
assumption, in fact, the final states from a pair of
$\tilde{\chi}^0_1$'s are:
\begin{equation}
 2(b \bar{b}) \,(2 \nu_\tau) \,,
\label{LSPpairone} 
\end{equation}
and 
\begin{equation}
 (2t)\,(2\bar{b}) \,(2\tau)\,,                \qquad 
 (2\bar{t})\,(2b) \,(2\bar{\tau})\,,          \qquad 
 t \bar{t} \,b \bar{b} \,\tau \bar{\tau}\,,   \qquad 
 t b \,(2\bar{b}) \,\tau \nu_\tau\,,          \qquad 
 \bar{t} \bar{b} \,(2b) \bar{\tau} \nu_\tau\,,  
\label{LSPpairtwo} 
\end{equation}
if the lightest neutralino is heavier than the $t$-quark.

It is clear, then, that the $3$-body decays of both,
$\tilde{\chi}^0_1$ and $\tilde{\chi}^-_1$, induced by large
$R_p$-violating terms are worth a detailed study. For masses of
$\tilde{\chi}^0_1$ and $\tilde{\chi}^-_1$ of interest to LHC and/or
Linear Collider studies, decays of $\tilde{\chi}^0_1$ and
$\tilde{\chi}^-_1$ into third generation fermions are likely to be
allowed kinematically.  Further, we expect the third generation
sfermions, which are exchanged as virtual particles in these
processes, to be lighter than the others. Effects of the mass of the
fermions produced in $3$-body decays of the charginos and heavier
neutralinos were considered in the MSSM~\cite{Bartl:1999iw} and in
mSUGRA~\cite{Baer:1998bj}, in the context of $R_p$-conserving models.
These show that $3$-body decays into final states containing third
generation fermions, such as $b\bar b, \tau^+ \tau^-$ ($\tau \nu$),
are dominant. In certain regions of the parameter space the
contributions due to the Higgs boson exchange are large even for
moderate values of $\tan\beta$.  It is interesting to find out if
these predictions remain valid in the context of $R_p$-violating
models. Moreover, it is also important to know whether the decays of
$\tilde{\chi}^0_1$ and $\tilde{\chi}^-_1$ into the heaviest third
generation quark, the $t$-quark, have a competitive rate, in spite of
the large kinematical suppression.

The $3$-body decays of the lightest neutralino in the presence of
$R_p$-violating couplings have been studied earlier and expressions
for the differential decay widths exist in the literature, see
refs.~\cite{Butterworth:1993tc}--\cite{Dreiner:2000qz}. Correct matrix
elements squared, however, were obtained first in
ref.~\cite{Baltz:1998gd}, up to a typographical error in the sign of
the width of the sfermions exchanged as virtual particles in these
decays.  This was corrected in ref.~\cite{Dreiner:2000qz}. No
phenomenological analysis is presented in~\cite{Baltz:1998gd}, whereas
the analysis in~\cite{Ghosh:1999ix} concentrates on one of the
signatures caused by these decays without addressing the issue of
relative decay widths and branching ratios.  The phenomenological
analysis in~\cite{Dreiner:2000qz} deals with the
$\lambda^{\prime\prime}$ couplings, although formulae for all
$R_p$-violating couplings are listed. The attitude in
ref.~\cite{Dreiner:2000qz} is opposite to that taken here: a solution
to the problem of a too fast proton decay is obtained by allowing only
the $B$-violating couplings instead than the $L$-violating ones.

The aim of this paper is to investigate the $3$-body decays of the
lightest neutralino into mainly third generation fermions, including
the $t$-quark, induced by the $R_p$-violating couplings
$\lambda^\prime$ and $\lambda$.  This is also a preliminary and
nontrivial step towards a complete study of the $3$-body decays of the
lightest chargino. If kinematically possible, this may decay into the
lightest neutralino through gauge interactions, i.e.\
$\tilde{\chi}^-_1 \to \tilde{\chi}^0_1 \bar{f}_u f_d $ and
$\tilde{\chi}^-_1 \to \tilde{\chi}^0_1 {\nu}_l l^-$. Such a study is
left for further work~\cite{PREP}.

This paper is organized as follows. After a discussion in
section~\ref{limits} of the experimental constraints on
$\lambda^\prime$ and $\lambda$ couplings, we list in
section~\ref{interactions} all the relevant interaction terms. In
section~\ref{lambdaprime333} are given the complete analytical
formulae for the $3$-body decays of ${\tilde\chi}_1^0$, whether it is
the LSP or not, in the case of a dominant $\lambda'_{333}$ coupling.
No assumption is made on the actual composition of ${\tilde\chi}_1^0$
(i.e.\ if pure Bino ($\tilde{B}$), or mixed Bino-Wino
($\tilde{B}$-$\tilde{W}$), or mixed gaugino and Higgsino
($\tilde{H}$)), and complex values of the left-right mixing parameters
in the sfermion mass matrices are assumed.  These formulae provide a
further check on the calculation of the widths of the $3$-body
neutralino decays, which has proven to be rather nontrivial. We
confirm the analytic results of ref.~\cite{Dreiner:2000qz}, when
taking the limit of real left-right mixing masses in the sfermion mass
matrices.  Our formulae can be easily generalized to study decays of
heavier neutralinos and to situations in which other $R_p$-violating
couplings are present. In particular, we use them to discuss the
decays induced by sizable values of $\lambda_{233}$ and
$\lambda^\prime_{233}$. Moreover, the generality in the composition of
${\tilde\chi}_1^0$, allows model-independent analyses. In spite of the
large number of parameters on which the numerical results depend, we
can capture some essential features of the different decay widths by
choosing different values of slepton/squark masses, for different
gaugino/higgsino contents of the neutralino, and by choosing small and
large left-right mixing terms in the sfermion sector.  We present and
discuss these features in section~\ref{results}. We end by summarizing
our results in section~\ref{concl}.  Appendices~\ref{neutralino}
and~\ref{sfermion} give our conventions for the neutralino and
sfermion mass matrices; appendix~\ref{sf_widths} gives the expressions
for the $R_p$-violating $2$-body decay widths of squarks and sleptons
when $\lambda^\prime_{333}$ and $\lambda_{233}$ are nonvanishing,
whereas appendix~\ref{squaredoperators} gives the expressions of terms
required to evaluate the matrix element squared for the decay of the
lightest neutralino.

\section{Limits on $R_p$-violating couplings}
\label{limits}

Hints on the size of the $R_p$-violating couplings in
eq.~(\ref{superpot}) may be obtained from various experiments. Two
types of experimental constraints are possible. There are ``direct''
constraints due to effects of these couplings on sparticle production
and on the decays of particle/sparticles at colliders.  ``Indirect''
constraints~\cite{BGH}, instead, are coming from measurements of SM
observables to which supersymmetric particles contribute as virtual
particles at the quantum level, or even at the tree level, as in
scattering processes.  Such observables include, at present, the EDM
of fermions~\cite{Godbole:2000ye}, the anomalous magnetic moment of
the muon~\cite{Kim:2001se}, neutrino masses, the decay widths of the
$Z$-~\cite{Bhattacharyya:1994ig,Lebedev:2000ze,Lebedev:2000vc} and of
the $W$-boson~\cite{Lebedev:2000vc} the strength of four-fermion
interactions, with the subsequent production of lepton pairs at
hadron~\cite{Hewett:1997ce,godbolechoudhury} and lepton
colliders~\cite{leptcolliders}, rare processes such as $\mu \to
e\gamma$~\cite{MUEGAMMA} and $b \to s \gamma$~\cite{RPBSGAMMA}, the
$e-\mu-\tau$
universality~\cite{Lebedev:2000vc,GDRrep,Allanach:1999ic}, etc.

Among these, a particular role is played by neutrino masses, which,
being very small, restrict quite strongly the size of the dimensionful
couplings $\kappa_i$ in eq.~(\ref{superpot}) and of the vacuum
expectation values (\emph{vev}'s) of the neutral scalar components of
the fields $L_i$, $v_i$~\cite{RPV-NUM-tree}.  Strictly speaking,
neutrino physics constrains only the parameter $(k_i\cdot v_i)/(k_i^2
+\mu^2)^{1/2} (v_i^2 +v_d^2)^{1/2}$.  If not very small, the vectors
$\{v_i\}$ and $\{k_i\}$ must be quite accurately aligned. In the
following, we assume, for simplicity, that both vectors have very
small magnitudes.  (Decays of the lightest neutralino induced by
bilinear couplings were studied in~\cite{CHI0BILINEAR}.)

Trilinear couplings such as $\lambda_{i33}$ and
$\lambda^\prime_{i33}$, with $i=1,2$, may also be severely bound by
the value of the Majorana mass terms for $\nu_e$ and $\nu_\mu$ that
they induce at the loop level~\cite{Godbole:1993fb}. The specific
value of these bounds depends on the mass of the third-generation
sfermion doublet and singlet as well as on the mass term that mixes
them. Most of all, it depends also on the particular neutrino spectrum
assumed, for example whether $\nu_e$ and $\nu_\mu$ are in the sub-eV
region, as the neutrino oscillation experiments may imply. If the
heaviest neutrino is also $\ltap 1\,$eV, then, a constraint may be
deduced also for $\lambda^\prime_{333}$ (but not for
$\lambda^\prime_{323}$ or $\lambda^\prime_{332}$), which induces a
quantum contribution to the mass of $\nu_\tau$, in addition to that
generated at the tree-level by the dimensionful couplings $\kappa_i$.
However, neutrino masses may be efficiently suppressed not only by
small $\lambda$ and $\lambda^\prime$ couplings, but also by very tiny
left-right mixing terms in the slepton and squark sectors, and/or
large values of the slepton and squark masses. Notice that very small
left-right sfermion mass terms, as well as large sfermion mass
eigenvalues, do change quantitatively the predictions for the widths
of neutralino and chargino decays, but do not spoil the possibility of
visible signals for these decays. Because of all this, and because the
neutrino spectrum is not yet known, we prefer to leave the constraints
from neutrino masses aside and consider them as complementary to those
that can be obtained from collider searches, in a direct way.
Nevertheless, it should be remembered that, irrespectively of
constraints from neutrino masses, at least one combination of the
$L$-violating couplings must be very small if leptogenesis has to
occur~\cite{Dreiner:1993vm,Ma:1999ni}.

In general, the ``direct'' and ``indirect'' observables listed above
constrain most of the $R_p$-violating couplings, in particular those
involving first and second generation indices, to small values,
smaller than the gauge
couplings~\cite{BGH,Allanach:1999ic,BOUNDS1}. These bounds depend on
the values of other supersymmetric parameters and, at the 2$\sigma$
level, they are severe for rather small soft scalar masses, i.e.\
$\sim 100\,$GeV~\cite{Allanach:1999ic}. The situation is somewhat
different for couplings involving third generation particles, which
are left practically unrestricted by the low energy processes
mentioned above. Constraints on $\lambda^\prime_{3jk}$ (and
$\lambda^{\prime\prime}_{3jk}$) come, for example, indirectly, from
radiative corrections to $Z \to b \bar b$, i.e.\ $\ltap
1$~\cite{Bhattacharyya:1994ig,Lebedev:2000ze,BOUNDS1}, and to $Z \to
l^+ l^-$, i.e $\ltap 0.42$~\cite{Lebedev:2000vc}. Both bounds are
derived for sfermion (slepton/squark) masses $\sim 100\,$GeV.  For
heavier sfermions, the constraint on $\lambda^\prime_{3jk}$ (and
$\lambda^{\prime \prime}_{3jk}$) are somewhat weaker.  Rare processes
like $b \to s \gamma$ involve far too many parameters to give
clear-cut constraints on any of them~\cite{RPBSGAMMA}.

In the future, constraints to the $R_p$-violating couplings involving
third-generation indices may be obtained indirectly from other
interesting quantities, such as the $t$-channel contribution to the
$t\bar t$ production, $t$-quark decays into SM
fermions~\cite{Ghosh:1997bm,Hikasa:1999wy,TOPDEC}, and the
polarization of the $t/\bar t$ quark, which may all get contributions
for nonvanishing $R_p$-violating couplings. In particular, studies of
the $t/\bar t$ quark polarization at the Tevatron can probe
$\lambda^\prime_{31i}$ (and $\lambda^{\prime \prime}_{311}$), to
somewhat lower values~\cite{Hikasa:1999wy,gdbolepoulose} than those
tested by the $Z$ width. Finally, the effect of a $\tilde{t}$-squark
exchange in the $t$ channel on the Drell-Yan $\mu^+ \mu^-$ pair
production, not very significant at the Tevatron
energy~\cite{Hewett:1997ce}, may probe $\lambda^\prime_{231}$, at the
LHC energies, to values $< 0.2$ even for $\tilde{t}$-squark masses as
large as $800\,$GeV~\cite{godbolechoudhury}.

Direct constraints on $R_p$-violating couplings that are comparable or
larger than the gauge couplings, i.e. $\gtap 10^{-1}$--$10^{-2}$,
are obtained from searches of: {\it i)} unusual decays of the
$t$-quark~\cite{TOPDEC} into still allowed light superpartners, {\it
ii)} resonances which violate $L$ or $B$, {\it iii)} signals due to
the decays of superpartners, among them the lightest neutralino.  The
last, in particular, may be the only signal to be probed, for
couplings smaller than $10^{-2}$. At a hadron collider,
$\lambda^\prime$ couplings can be probed by searching for resonant
slepton production~\cite{Moreau:2001bs}; $\lambda^\prime$ and
$\lambda$ couplings when both production and decay of such resonances
take place via $R_p$ violating couplings~\cite{Allanach:1999bf}. At a
lepton collider the $\lambda$ couplings~\cite{Erler:1997ww} are those
to be probed.

Searches at the Tevatron rule out a first generation squark up to
$100\,$GeV, almost irrespective of the size of the $\lambda^\prime$
couplings~\cite{TEVREFS}. These direct searches put limits on the
$\lambda$ and $\lambda^\prime$ couplings involving mainly the first
two generations, as both the colliding particles as well as the final
state particles are in general from the first two
generations~\cite{TEVREFS2}.

At the $ep$ collider HERA a squark resonance can be produced via
interactions with $\lambda^\prime$ couplings. The nonobservation of
such a resonance in the $e^+p$ data has ruled out a first generation
squark up to $260\,$GeV for a coupling (to a quark and a lepton)
$\lambda^\prime \sim \sqrt{4 \pi \alpha_{em}}$~\cite{H1}; $e^-p$ data
should complement this analysis~\cite{H1n} and strengthen this bound.
The analysis for the HERA data has been done in the framework of an
unconstrained MSSM, with the additional assumption of a common gaugino
mass at a high scale, as well as in the mSUGRA framework.

Searches of s-channel resonance formation at LEP have yielded
constraints on the couplings $\lambda_{121}$, $\lambda_{131}$ and
$\lambda_{232}$ in a model independent
way~\cite{l3old,opalnew,alephnew}.  (These analyses include also
indirect effects of the t-channel sfermion exchange on the four
fermion scattering.)  These constraints are considerable for low
values of sneutrino masses, viz.\  $100 < m_{\tilde \nu} < 200$
GeV, but rise close to the values of gauge couplings for higher
sneutrino masses~\cite{l3old,opalnew,alephnew}.  In general, for
sfermion masses $\gtap 100 $\,GeV, LEP experiments do not give
substantial limits on any of the R-parity violating couplings
involving more than one third generation index.  LEP has also probed
R-parity violating couplings, by searching for decays of sparticles,
including the decay of the lightest neutralino into light fermions
(see for example ref.~\cite{Godbole:1993fb}).  These searches have
essentially yielded only limits on sparticle masses~\cite{opal_aleph}.

Some of these analyses, have been carried out in the framework of
mSUGRA, without keeping into account the fact that the effect of
relatively large $R_p$-violating couplings cannot be ignored in the
determination of the mass spectra.

In view of all the above considerations, in the following analysis we
use the values $0.5$, $1$ for $\lambda^\prime_{333}$ and
$\lambda^\prime_{233}$, whereas we choose a somewhat lower value,
i.e.\ $0.2$, for $\lambda_{233}$.  For one particular direction of
parameter space, we show how the decay widths change when the value of
the relevant $R_p$-violating couplings are lowered.  In the regions
where the sfermions mediating these decays are nonresonant, the decay
widths are simply scaled down by the overall factors
$\vert\lambda^\prime_{i33}\vert^2$ or
$\vert\lambda_{233}\vert^2$. They remain, however, marginally affected
in the sfermion resonant regions.  This behaviour can be extrapolated
to the other directions of parameter space studied in this paper. The
present analysis therefore applies also to the case in which the
relevant couplings are small, provided they dominate over other
$R_p$-violating couplings.

\section{Interaction terms relevant for $\tilde{\chi}^0_1$ decays}
\label{interactions} 

The interactions relevant for an analysis of neutralino decays are:
\emph{i)} the $R_p$-violating interactions due to the first two terms
of the superpotential in eq.~(\ref{superpot}), and \emph{ii)} the
neutralino-fermion-sfermion couplings due to gauge and Yukawa
interactions.

In the following, we give explicitly the fermion-fermion-sfermion
interaction terms derived from $\lambda_{ijk} L_i L_j E^c_k $ and
$\lambda_{ijk}^\prime L_i Q_j D^c_k $.  Assuming the standard
contraction of SU(2) indices:
\begin{equation}
\lambda_{ijk}^\prime L_i Q_j D^c_k = \lambda_{ijk}^\prime
\epsilon_{ab} \left(L_i\right)_a \left(Q_j\right)_b D^c_k\,,
\label{sutwocontraction}
\end{equation}
where $\epsilon_{12} = -\epsilon_{21} = 1$, the $R_p$-violating 
interaction terms with couplings $\lambda^\prime_{ijk}$, are:
\begin{eqnarray}
 {\cal L} & = & + \lambda_{ijk}^\prime \, \tilde{u}_{jL}
\left(\overline{d_{kR}}\,l_{iL} \right) + \lambda_{ijk}^\prime \,
\tilde{l}_{iL} \left(\overline{d_{kR}}\, u_{jL} \right) +
\lambda_{ijk}^\prime \, \tilde{d}_{kR}^\ast
\left(\overline{{l^c}_{iR}}\, u_{jL} \right) + 
\nonumber \\ & & 
{}+ \lambda_{ijk}^{\prime \ast}\, \tilde{u}_{jL}^\ast
\left(\overline{l_{iL}}\, d_{kR} \right) + \lambda_{ijk}^{\prime
\ast}\, \tilde{l}_{iL}^\ast \left(\overline{u_{jL}}\, d_{kR} \right) +
\lambda_{ijk}^{\prime \ast}\, \tilde{d}_{kR} \left(\overline{u_{jL}}\,
{l^c}_{iR}\right),
\label{int-lambdap-ch}
\end{eqnarray}
when the charged lepton $l$ is involved, and 
\begin{eqnarray}
{\cal L} & = & - \lambda_{ijk}^\prime \, \tilde{d}_{jL}
\left(\overline{d_{kR}}\,\nu_{iL} \right) - \lambda_{ijk}^\prime \,
\tilde{\nu}_{iL} \left(\overline{d_{kR}}\, d_{jL} \right) -
\lambda_{ijk}^\prime \, \tilde{d}_{kR}^\ast
\left(\overline{{\nu^c}_{iR}}\, d_{jL} \right) -
\nonumber \\  &   &  
{}- \lambda_{ijk}^{\prime \ast} \, \tilde{d}_{jL}^\ast
\left(\overline{\nu_{iL}}\, d_{kR} \right) - \lambda_{ijk}^{\prime
\ast} \, \tilde{\nu}_{iL}^\ast \left(\overline{d_{jL}}\, d_{kR}
\right) - \lambda_{ijk}^{\prime \ast} \, \tilde{d}_{kR}
\left(\overline{d_{jL}}\, {\nu^c}_{iR} \right),
\label{int-lambdap-neu}
\end{eqnarray}
when the lepton interacting is neutral. The upperscript $^c$ 
indicates charge conjugation ($\psi^c = C \bar{\psi}^T$). 

The interaction terms with couplings of type $\lambda_{ijk}$ can be
obtained in a similar way:
\begin{eqnarray}
 {\cal L}  & = &
+  \lambda_{ijk}    \, 
\tilde{\nu}_{jL}    \left(\overline{l_{kR}}\, l_{iL}       \right)
+  \lambda_{ijk}    \,
\tilde{l}_{iL}      \left(\overline{l_{kR}}\, \nu_{jL}     \right)
+  \lambda_{ijk}    \, 
\tilde{l}_{kR}^\ast \left(\overline{{l^c}_{iR}}\, \nu_{jL} \right)  + 
\nonumber \\
           &   &  
{}+  \lambda_{ijk}^{\ast}\, 
\tilde{\nu}_{jL}^\ast  \left(\overline{l_{iL}}\, l_{kR}      \right)
+  \lambda_{ijk}^{\ast}\,
\tilde{l}_{iL}^\ast    \left(\overline{\nu_{jL}}\, l_{kR}    \right)
+  \lambda_{ijk}^{\ast}\, 
\tilde{l}_{kR}         \left(\overline{\nu_{jL}}\, {l^c}_{iR}\right) \,,
\qquad(i>j)  \qquad
\label{int-lambda}
\end{eqnarray}
where the relation $\lambda_{ijk}=-\lambda_{jik}$ was used. Notice 
that there are only 9 independent $\lambda$ couplings versus the 27 
of type $\lambda^\prime$. 

The neutralino eigenstates are obtained diagonalizing the matrix shown
in appendix~\ref{neutralino}. Strictly speaking, in $R_p$-violating
models the neutralino mass matrix is a $7 \times 7$ matrix. In the
approximation of very small bilinear couplings $k_i$ and \emph{vev}'s
$v_i$ of the neutral components of the fields $\tilde{L}_i$, this
reduces to the conventional $4\times 4$ matrix typical of
$R_p$-conserving models. Thus, the neutralino-sfermion-fermion
interaction terms are:
\begin{eqnarray}
L  & = & -\sqrt{2}g\sum_i
\biggl\{
 L_i^{f_L}\left( \overline{({\tilde{\chi}^0}_{iR})}\cdot f_L
          \right)\tilde{f}_L^{\ast}
-R_i^{f_L}\left( \overline{({\tilde{\chi}^0}_{iL})}\cdot f_R
          \right)\tilde{f}_L^{\ast}
 + 
\nonumber \\& & 
\hphantom{-\sqrt{2}g\sum_i\biggl\{}
+ L_i^{f_R}\left( \overline{({\tilde{\chi}^0}_{iR})}\cdot f_L
          \right)\tilde{f}_R^{\ast}
-R_i^{f_R}\left( \overline{({\tilde{\chi}^0}_{iL})}\cdot f_R
          \right)\tilde{f}_R^{\ast}
\biggr\}  +  {\rm h.c.} 
\label{neut-ferm-sferm}
\end{eqnarray}
where $i$ is now a neutralino index $(i=1,2,3,4)$ and $g$ is the
SU(2) gauge coupling. By using the definition of hypercharge 
$ Q =(Y+T_3)$, the coefficients $L_i^{f_{L}}$ and $R_i^{f_{R}}$ can be
cast in the compact form:
\begin{equation}
 L_i^{f_L}  = \eta_i^{\ast}
   \left( T_{3\,f_L} O_{i2}+ Y_{f_L} O_{i1}\tan\theta_W\right),
\qquad 
 R_i^{f_R}  = \eta_i \, Y_{f_R} O_{i1}\tan\theta_W \,,
\label{gaugecouplings}
\end{equation}
whereas the coefficients $L_i^{f_{R}}$ and $R_i^{f_{L}}$ are different
for fermions of up ($f_u$) and down type~($f_d$):
\begin{eqnarray}
 L_i^{(f_u)_R} = +\frac{1}{2} \,\eta_i^{\ast}\, 
                 \frac{m_{f_u}}{M_W \sin \beta} \, O_{i4} \,,
\quad & & \quad  
 R_i^{(f_u)_L} = -\frac{1}{2} \,\eta_i \,
                 \frac{m_{f_u}}{M_W \sin \beta} \, O_{i4} \,,
\nonumber \\[1.001ex]
 L_i^{(f_d)_R} = +\frac{1}{2} \,\eta_i^{\ast}\, 
                 \frac{m_{f_d}}{M_W \cos \beta} \, O_{i3} \,,
\quad & & \quad 
 R_i^{(f_d)_L} = -\frac{1}{2} \,\eta_i \,
                 \frac{m_{f_d}}{M_W \cos \beta} \, O_{i3} \,.
\label{higgscouplings}
\end{eqnarray}
Notice that all neutrino couplings are vanishing except for
$L_i^{\nu_L}$.  The matrix $O$ in the previous equations is the
neutralino diagonalization matrix.  The phase $\eta$ is due to the
fact that the diagonalization of the neutralino mass matrix may yield
complex eigenvalues, see appendix~\ref{neutralino}. The neutralino
eigenstates may be rotated in such a way to obtain real and positive
eigenvalues, but phase factors $\eta$ appear in this case in the
interaction terms.  For real values of the parameters $M_{\tilde B}$,
$M_{\tilde W}$, and $\mu$, the mass eigenvalues are real, but with
signs that can be positive or negative. A rotation of the eigenstates
that reduces all eigenvalues to be positive, brings in factors
$\eta_i$, which are now $i$ or $1$.

In the approximation of pure $B$-ino for the lightest neutralino, the
coefficients $L_i^{f_{R}}$ and $R_i^{f_{L}}$ vanish identically, those
in eq.~(\ref{gaugecouplings}) become $ L_i^{f_L} = \eta_i^{\ast}
Y_{f_L} \tan\theta_W$, $ R_i^{f_R} = \eta_i Y_{f_R} \tan\theta_W$.
Equation~(\ref{neut-ferm-sferm}) has, then, the simple form:
\begin{equation}
{\cal L} = -\sqrt{2}\,g_Y \left\{ \eta_1^\ast Y_{f_L}
\left(\overline{\tilde{\chi}^0_{1R}}\,f_L\right)\tilde{f}_L^{\ast} -
 \eta_1 Y_{f_R}
\left(\overline{\tilde{\chi}^0_{1L}}\,f_R\right)\tilde{f}_R^{\ast}
\right\}  + {\rm h.c}
\label{bino-ferm-sferm}
\end{equation}
where $g_Y$ is expressed in terms of $g$ as $ g_Y = g
\tan\theta_W$. In this approximation, unless the gaugino mass $M_1$ is
chosen to be negative, the phase factor $\eta$ can be altogether
dropped.

In the approximation of pure $B$-ino, the lightest neutralino decays
via gauge interactions into a fermion-sfermion pair, $\bar{f}
\tilde{f}$. The sfermion $\bar{f} \tilde{f}$ is off-shell when
$\tilde{\chi}^0_{1}$ is the LSP. This sfermion decays predominantly
through $R_p$-violating interactions, if the corresponding couplings
are large.

In the following, we start considering the case of a dominant
$\lambda_{333}^\prime$ coupling and discuss later the possibility of
having simultaneously two relatively large couplings, for example
$\lambda_{333}^\prime$ and $\lambda_{233}^\prime$, or
$\lambda_{333}^\prime$ and $\lambda_{233}$.

\section{Dominant $\lambda_{333}^\prime$ coupling}  
\label{lambdaprime333}

If $\lambda_{333}^\prime$ is the dominant $R_p$-violating coupling,
the only relevant decays of the lightest neutralino into a
fermion-sfermion pair are:
\begin{equation}
 \bar{b}         \tilde{b}         \,,  \qquad  
 \bar{\nu}_\tau  \tilde{\nu}_\tau  \,,  \qquad 
 \bar{\tau}      \tilde{\tau}      \,,
\label{gdec_lightneutr} 
\end{equation}
and their CP-conjugated ones, for a lightest neutralino lighter than the
top-quark, or into:
\begin{equation}
 \bar{b}         \tilde{b}         \,,  \qquad  
 \bar{\nu}_\tau  \tilde{\nu}_\tau  \,,  \qquad 
 \bar{\tau}      \tilde{\tau}      \,,  \qquad
 \bar{t}         \tilde{t}         \,, 
\label{gdec_heavyneutr} 
\end{equation}
again with their CP-conjugated ones, for a lightest neutralino heavier
than the top-quark. In the first case, the two decay modes
$\bar{b}\tilde{b}$ and $\bar{\nu}_\tau \tilde{\nu}_\tau$ give rise to
the final state $\bar{b} b \nu_\tau$.  Since neutrinos can be safely
considered as massless at the energies at which neutralino decays will
be studied, in the following, the two possible final states $\bar{b} b
\nu_\tau$ and $\bar{b} b \bar{\nu}_\tau$ are identified.  The third
mode in eq.~(\ref{gdec_lightneutr}), $\bar{\tau}\tilde{\tau}$, gives
rise to an off-shell top-quark through the decay $\tilde{\tau} \to b
\bar{t}$ (see eq.~(\ref{staudecRP})) and therefore a multi-body final
state. If $\tilde{\chi}^0_{1}$ is heavier than the top-quark, the
modes $\bar{\tau} \tilde{\tau}$ and $\bar{t} \tilde{t}$ give rise to
the final state $\bar{t} b \bar{\tau}$, to the conjugate of which,
also the mode $ \bar{b}\tilde{b}$ may contribute.  Therefore, in the
above approximation of only one $R_p$-violating coupling, the $3$-body
final states arising from the decay of $\tilde{\chi}^0_{1}$ are an
almost massless one, $\bar{b} b \nu_\tau$ and a massive one, $\bar{b}
t \tau$.  In the following, we give amplitudes and widths for both
decays.

\subsection{Massless final state --- $\tilde{\chi}^0_1\to b
  \bar{b}\nu_\tau$}  
\label{massless} 

\begin{figure}[ht] 
\begin{center}
\epsfxsize= 12 cm 
\leavevmode 
\epsfbox{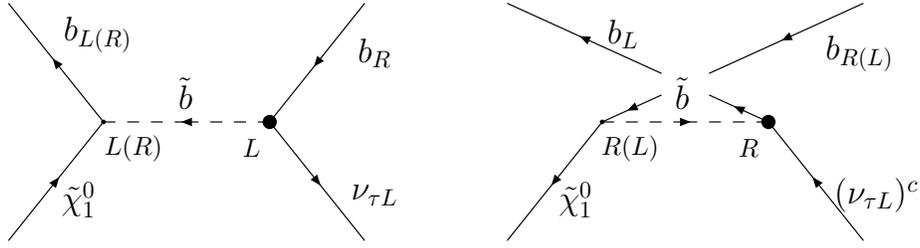}
\end{center} 
\caption{Diagrams
contributing to the decay $\tilde{\chi}^0_1\to b \bar{b}\nu_\tau$
through the exchange of a virtual $\tilde{b}$-squark. The thick vertex
indicates the $R_p$-violating coupling
$\lambda_{333}^{\prime\ast}$. The labels $L$ and $R$ at each vertex
indicate the chirality of the ingoing and/or outgoing scalar
fields. To the same decay contribute also the crossed diagrams, i.e.\
those with the two external bottom-quarks interchanged and an ingoing
neutrino line, and with the thick vertex indicating the
$R_p$-violating coupling $\lambda_{333}^{\prime}$.}\label{diagsbot}
\end{figure}

\begin{figure}[ht] 
\begin{center}
\epsfxsize= 3.4 cm 
\leavevmode 
\epsfbox{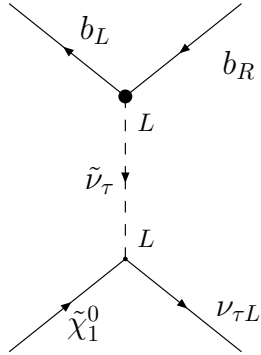}
\end{center} 
\caption{Diagram contributing to the decay
$\tilde{\chi}^0_1\to b \bar{b} \nu_\tau $, in particular to
$\tilde{\chi}^0_1\to b_L \overline{(b_R)}\, \nu_{\tau L}$, through the
exchange of a virtual neutral slepton $\tilde{\nu}_\tau$. The thick
vertex indicates the $R_p$-violating coupling
$\lambda_{333}^{\prime\ast}$. The label $L$ at each vertex indicate
the chirality of the ingoing/outgoing scalar field.  A similar
diagram, with the thick vertex indicating $\lambda_{333}^{\prime}$,
contributes to the decay $\tilde{\chi}^0_1\to b_R \overline{(b_L)}\,
\overline{\nu_{\tau L}}$.}\label{diagsneu}
\end{figure}

The decay $\tilde{\chi}^0_1\to b \bar{b}\nu_\tau$ is mediated by the
exchange of an off-shell/on-shell $\tilde{b}$-squark, as in the
diagrams of figure~\ref{diagsbot}, and an off-shell/on-shell neutral
slepton $\tilde{\nu}_\tau$, as shown by the diagram in
figure~\ref{diagsneu}. In both figures, only the diagrams giving rise
to matrix elements proportional to $\lambda_{333}^{\prime\,\ast}$ are
shown. To the same decay contribute also the ``crossed'' diagrams,
i.e.\ those with the two external $b$-quarks interchanged and an
ingoing neutrino line.  The corresponding matrix elements are
proportional to $\lambda_{333}^{\prime}$.

It is convenient to split the matrix elements of all contributions to
both decays $\tilde{\chi}^0_1\to b \bar{b}\nu_\tau$ and
$\tilde{\chi}^0_1\to t \bar{b} \tau$ as follows:
\begin{equation}
 M =  \sum_{s,t} a^t_{S,s} \, Q^t_{S,s}\,.  
\label{ampl_decomp} 
\end{equation}
The $Q$ terms are the results of the contractions between initial and
final states of all possible operators obtained from the lagrangian
interaction terms, taken in absolute value and without numerical
coefficients.  Possible relative minus signs, which may be obtained in
these contractions according to the Wick's theorem, are included in
the corresponding coefficients.

The $Q$ terms contributing to the decay $\tilde{\chi}^0_1(q)\to
b(p_1)\nu_\tau(p_2) \bar{b}(p_3)$,
\begin{eqnarray}
 Q_{S,RR}^x &=& 
 \left(\overline{u}_{\nu_\tau}(p_2) P_R\,v_{b}   (p_3)\right)
 \left(\overline{u}_{b}  (p_1) P_R\,u_{{\tilde{\chi}}^0_1}(q)\right)
\nonumber  \\
 Q_{S,RL}^x &=& 
 \left(\overline{u}_{\nu_\tau}(p_2) P_R\,v_{b}   (p_3)\right)
 \left(\overline{u}_{b}  (p_1) P_L\,u_{{\tilde{\chi}}^0_1}(q)\right)
\nonumber  \\
 Q_{S,RR}^z &=& 
 \left(\overline{u}_{\nu_\tau}(p_2) P_R\,v_{b}   (p_1)\right)
 \left(\overline{u}_{b}  (p_3) P_R\,u_{{\tilde{\chi}}^0_1}(q)\right)
\nonumber  \\
 Q_{S,RL}^z &=& 
 \left(\overline{u}_{\nu_\tau}(p_2) P_R\,v_{b}   (p_1)\right)
 \left(\overline{u}_{b}  (p_3) P_L\,u_{{\tilde{\chi}}^0_1}(q)\right)
\nonumber  \\
 Q_{S,RR}^y &=& 
 \left(\overline{u}_{b}  (p_1) P_R\,v_{b}   (p_3)\right)
 \left(\overline{u}_{\nu_\tau}(p_2)
 P_R\,u_{{\tilde{\chi}}^0_1}(q)\right) . 
\label{q_diagr}
\end{eqnarray}
have as corresponding coefficients:
\begin{equation}
\begin{array}{lllll}
a_{S,RR}^{x}  &     =          & -\sqrt{2}\,g 
 \lambda_{333}^{\prime \ast}
 \left((L^{b_L}_1)^{\ast}D_{LL}^{\tilde{b}}(x)
      +(L^{b_R}_1)^{\ast}D_{RL}^{\tilde{b}}(x)\right)     
              &     \to       & -\sqrt{2}\,g_Y
 \lambda_{333}^{\prime \ast}\,\eta_1^{\ast} \,Y_{b_L}     
 D_{LL}^{\tilde{b}}(x)
\nonumber\\[1ex]
a_{S,RL}^{x}  &     =          & +\sqrt{2}\,g 
 \lambda_{333}^{\prime \ast}
 \left((R^{b_R}_1)^{\ast} D_{RL}^{\tilde{b}}(x)
      +(R^{b_L}_1)^{\ast} D_{LL}^{\tilde{b}}(x)
\right)     
              &     \to       & +\sqrt{2}\,g_Y
 \lambda_{333}^{\prime \ast}\,\eta_1^{\ast}\,Y_{b_R}
 D_{RL}^{\tilde{b}}(x)
\nonumber\\[1ex]
a_{S,RR}^{z}  &     =          & -\sqrt{2}\,g 
 \lambda_{333}^{\prime \ast}
 \left(R^{b_R}_1(D_{RR}^{\tilde{b}}(z))^{\ast}
      +R^{b_L}_1(D_{LR}^{\tilde{b}}(z))^{\ast}\right)\,          
              &     \to       & -\sqrt{2}\,g_Y 
 \lambda_{333}^{\prime \ast}\,\eta_1 \,Y_{b_R}     
 (D_{RR}^{\tilde{b}}(z))^{\ast}
\nonumber\\[1ex]
a_{S,RL}^{z}  &     =          & +\sqrt{2}\,g 
 \lambda_{333}^{\prime \ast}
 \left(L^{b_L}_1(D_{LR}^{\tilde{b}}(z))^{\ast}
      +L^{b_R}_1(D_{RR}^{\tilde{b}}(z))^{\ast}
 \right)\,          
              &     \to       & +\sqrt{2}\,g_Y 
 \lambda_{333}^{\prime \ast}\,\eta_1\,Y_{b_L}
 (D_{LR}^{\tilde{b}}(z))^{\ast}
\nonumber\\[1ex]
a_{S,RR}^{y}  &     =          & +\sqrt{2}\,g 
 \lambda_{333}^{\prime \ast}\left(
   (L^{\nu_{\tau L}}_1)^{\ast}D_{LL}^{\tilde{\nu}_\tau}(y)
 \right)   
              &     \to       & +\sqrt{2}\,g_Y 
 \lambda_{333}^{\prime \ast}\,\eta_1^{\ast}  \,Y_{\nu_{\tau L}}  
 D_{LL}^{\tilde{\nu}_\tau}(y)                     \,, 
\end{array}
\label{coeff_diagr}
\end{equation}
where the limiting values in the approximation of lightest neutralino
as pure $B$-ino were given. In both definitions, the indices
$\{S,RR\}$ and $\{S,RL\}$ are reminders of the fact that these matrix
elements are obtained from a product of two scalar currents, each of
them with a chirality projector specified by $R$/$L$.  The upper index
$x$, $y$, and $z$ indicates the fraction of the initial momentum
flowing through the scalar propagator mediating these diagrams. They
are respectively:
\begin{equation}
  x = \frac{(p_1-q)^2}{q^2}\,, \qquad 
  y = \frac{(p_2-q)^2}{q^2}\,, \qquad 
  z = \frac{(p_3-q)^2}{q^2}\,.
\label{xyzdefs}
\end{equation} 
In the decomposition of eq.~(\ref{ampl_decomp}), therefore, $s$ runs
over $RR$ and $RL$, in the terms induced by the diagrams in
figures~\ref{diagsbot} and~\ref{diagsneu}, and over $LL$ and $LR$ in
those induced by the ``crossed'' diagrams, and $t$ may be $x$, $y$,
and $z$.  Finally, the coefficients $D_{\sigma}^{\tilde{f}}(t)$
collect the contributions from the scalar propagators of the two
states $\tilde{f}_{1,2}$ weighted by the factors projecting states of
definite chirality into mass eigenstates.  (No intergenerational
mixing among sfermions is assumed here.) The explicit expression of
the two mass eigenvalues $m^2_{\tilde{f}_{1,2}}$ and projection
factors $\sin \theta_f$ and $\cos \theta_f$ in terms of the parameters
in the sfermion mass matrix are given in appendix~\ref{sfermion}.  The
index $\sigma$, which runs over $LL,LR,RL,RR$, indicates the chirality
of the ingoing and outgoing scalar fields at each vertex of the
diagrams in figures~\ref{diagsbot} and~\ref{diagsneu}. For
definiteness, for $t= x,y,z$, it is:
\begin{eqnarray}
  D_{LL}^{\tilde{f}}(t) &=&           
   \sin^2\theta_f D_{\tilde{f}_1}(t) + 
   \cos^2\theta_f D_{\tilde{f}_2}(t)
\nonumber\\
  D_{LR}^{\tilde{f}}(t) &=&           
   \sin  \theta_f \cos\theta_f e^{-i\phi_f} \left(
    D_{\tilde{f}_1}(t) -D_{\tilde{f}_2}(t)  \right)
\nonumber\\
  D_{RL}^{\tilde{f}}(t) &=&            
   \sin  \theta_f \cos\theta_f e^{+i\phi_f} \left(
    D_{\tilde{f}_1}(t) -D_{\tilde{f}_2}(t)  \right)
\nonumber\\
  D_{RR}^{\tilde{f}}(t) &=&           
   \cos^2\theta_f D_{\tilde{f}_1}(t) +
   \sin^2\theta_f D_{\tilde{f}_2}(t) \,,
\end{eqnarray}
with  
\begin{equation}
  D_{\tilde{f}_{1,2}}(t) = \frac{1}{m_{\tilde{\chi}^0_1}^2}
    \left(t -\frac{ m^{2}_{\tilde{f}_{1,2}}}{ m_{\tilde{\chi}^0_1}^2}
    \right)^{-1},
\end{equation} 
for off-shell sfermions $\tilde{f}_{1,2}$, or 
\begin{equation}
  D_{\tilde{f}_{i}}(t) = \frac{1}{m_{\tilde{\chi}^0_1}^2}
    \left(t -\frac{ m^{2}_{\tilde{f}_{i}}}{ m_{\tilde{\chi}^0_1}^2}
           +i\frac{ \Gamma_{\tilde{f}_{i}}  m_{\tilde{f}_{i}}}
                                          { m_{\tilde{\chi}^0_1}^2} 
    \right)^{-1},   
\label{widthpropagator}
\end{equation} 
in the case in which the sfermion $\tilde{f}_i$ is on shell.  The
width of possible sfermions lighter than the lightest neutralino
${\tilde{\chi}^0_1}$ will be discussed in
appendix~\ref{sf_widths}. Finally, the angle $\phi_f$ is the argument
of the left-right mixing term in the mass matrix of the sfermion
$\tilde{f}$, see appendix~\ref{sfermion}.

Similarly, the $Q$ terms for the crossed diagrams are:
\begin{eqnarray}
Q_{S,LL}^x &\equiv&
 \left(\overline{u}_{\nu_\tau}(p_2) P_L v_{b}   (p_3)\right)
 \left(\overline{u}_{b}  (p_1) P_L\,u_{{\tilde{\chi}}^0_1}(q)\right)
\nonumber  \\
Q_{S,LR}^x &\equiv& 
 \left(\overline{u}_{\nu_\tau}(p_2) P_L v_{b}   (p_3)\right)
 \left(\overline{u}_{b}  (p_1) P_R\,u_{{\tilde{\chi}}^0_1}(q)\right)
\nonumber  \\
Q_{S,LL}^z &\equiv& 
 \left(\overline{u}_{\nu_\tau}(p_2) P_L v_{b}   (p_1)\right)
 \left(\overline{u}_{b}  (p_3) P_L\,u_{{\tilde{\chi}}^0_1}(q)\right)
\nonumber  \\
Q_{S,LR}^z &\equiv& 
 \left(\overline{u}_{\nu_\tau}(p_2) P_L v_{b}   (p_1)\right)
 \left(\overline{u}_{b}  (p_3) P_R\,u_{{\tilde{\chi}}^0_1}(q)\right)
\nonumber  \\
Q_{S,LL}^y &\equiv& 
 \left(\overline{u}_{b}  (p_1) P_L v_{b}   (p_3)\right)
 \left(\overline{u}_{\nu_\tau}(p_2) P_L\,u_{{\tilde{\chi}}^0_1}(q)\right).
\label{q_crossdiagr}
\end{eqnarray}
They differ from those in eq.~(\ref{q_diagr}) by an exchange 
$R \leftrightarrow L$. The related coefficients, with their limiting
values in the approximation of lightest neutralino as pure $B$-ino,
are:
{\setlength{\arraycolsep}{.5pt}
\begin{equation}
\begin{array}{lllll}
a_{S,LL}^{x}  &   =         & +\sqrt{2}\,g 
 \lambda_{333}^{\prime}
 \left((R^{b_R}_1)^{ \ast}D_{RR}^{\tilde{b}}(x)
      +(R^{b_L}_1)^{ \ast}D_{LR}^{\tilde{b}}(x) \right)
              &   \to       & +\sqrt{2}\,g_Y
 \lambda_{333}^{\prime}\,\eta_1^{\ast}\,Y_{b_R} 
 \left(D_{RR}^{\tilde{b}}(x)\right)
\nonumber\\[1.01ex]   
a_{S,LR}^{x}  &   =         & -\sqrt{2}\,g 
 \lambda_{333}^{\prime}
 \left((L^{b_L}_1)^{ \ast}D_{LR}^{\tilde{b}}(x)
      +(L^{b_R}_1)^{ \ast}D_{RR}^{\tilde{b}}(x) \right)
              &   \to       & -\sqrt{2}\,g_Y
 \lambda_{333}^{\prime}\,\eta_1^{\ast} \,Y_{b_L}       
 \left(D_{LR}^{\tilde{b}}(x)\right)
\nonumber\\[1.01ex]
a_{S,LL}^{z}  &   =         & +\sqrt{2}\,g
 \lambda_{333}^{\prime}
 \left((L^{b_L}_1) (D_{LL}^{\tilde{b}}(z))^{\ast}
      +(L^{b_R}_1) (D_{RL}^{\tilde{b}}(z))^{\ast} \right)
              &   \to       & +\sqrt{2}\,g_Y
 \lambda_{333}^{\prime}\,\eta_1\,Y_{b_L} 
 \left(D_{LL}^{\tilde{b}}(z)\right)^{\ast}
\nonumber\\[1.01ex]
a_{S,LR}^{z}  &   =         & -\sqrt{2}\,g
 \lambda_{333}^{\prime}
 \left((R^{b_R}_1)(D_{RL}^{\tilde{b}}(z))^{\ast}
      +(R^{b_L}_1)(D_{LL}^{\tilde{b}}(z))^{\ast} \right)
              &   \to       & -\sqrt{2}\,g_Y
 \lambda_{333}^{\prime}\,\eta_1 \,Y_{b_R}       
 \left(D_{RL}^{\tilde{b}}(z)\right)^{\ast}
\nonumber\\[1.01ex] 
a_{S,LL}^{y}  &    =         & +\sqrt{2}\,g
 \lambda_{333}^{\prime}\,
 \left((L^{\nu_{\tau L}}_1)^{\ast}D_{LL}^{\tilde{\nu}_\tau}(y) \right)
              &    \to       & +\sqrt{2}\,g_Y
 \lambda_{333}^{\prime}\,\eta_1^{\ast},Y_{\nu_{\tau L}}\!
 \left(D_{LL}^{\tilde{\nu}_\tau}(y)\right).
\end{array}
\label{coeff_crossdiagr}
\end{equation}}
The results in eqs.~(\ref{coeff_diagr}) and~(\ref{coeff_crossdiagr})
coincide with those in ref.~\cite{Dreiner:2000qz}, up to 
charge conjugations. 

The width for the decay mode $\tilde{\chi}^0_1\to b \bar{b}\nu_\tau$ is
finally obtained after integration of the differential one:
\begin{equation}
 \frac{d\Gamma(\tilde{\chi}^0_1\to b\bar{b}\nu_\tau)}{dxdy} = 
 \frac{3\, m_{\tilde{\chi}^0_1}\,} {512\pi^3}  \, 
 \left\vert M \right\vert_{z=1-x-y}^2  \,,
\label{diff_width} 
\end{equation} 
given by the standard $3$-body phase-space factor multiplied by the
square of the matrix elements averaged over the neutralino spin and
summed over spin and color of the final fermions. The sum over all
spin configurations is included in $\left\vert M \right\vert^2$,
whereas the average over the neutralino spin and the sum over color
give an overall factor $3/2$ included in the numerical coefficient in
eq.~(\ref{diff_width}). The square $\left\vert M \right\vert^2$ can be
expressed in terms of the products
$\beta_{s,s^{\prime}}^{t,t^{\prime}}\equiv Q_{S,s}^t
Q_{S,s^{\prime}}^{t^{\prime}\,\dagger}$ given explicitly in
appendix~\ref{squaredoperators}. They are evaluated under the
assumption that the particle with momentum $p_1$ has a nonnegligible
mass $m_1$, with $m_1^2 = r m_{\tilde{\chi}^0_1}^2$. The expressions
relevant to the case of the decay $\tilde{\chi}^0_1\to b
\bar{b}\nu_\tau$ are, then, obtained by taking the limit $r \to 0$ in
eqs.~(\ref{spinsumoper}),~(\ref{spinsumoper_mixt})--(\ref{spinsumoper_mixst}),
and~(\ref{spinsumoper_mixnew}).  For comparisons with
ref.~\cite{Baltz:1998gd}, one should keep in mind that the formulae
listed above combine the partial widths for the decays into $b \bar{b}
\nu_\tau$ and into $b \bar{b} \bar{\nu}_\tau$, whereas only the
partial width for $\tilde{\chi}^0_1\to b\bar{b}\nu_\tau $ is
explicitly given in ref.~\cite{Baltz:1998gd}.  The formal expression
in eq.~(\ref{diff_width}) is used also to obtain the partial width of
all other decays discussed in this paper with the obvious
modifications: \emph{1)} $r$ is nonvanishing in the expression for
$\vert M\vert^2$ when decays into massive final states are considered;
\emph{2)} the overall factor 3 of color must be removed in the case of
decays into purely leptonic final states.

The total width is obtained integrating eq.~(\ref{diff_width}) over
the two variables $x$ and $y$, with integration bounds given by:
\begin{equation}
  r \le  y  \le 1                      \,, \qquad 
  0 \le  x  \le (1-y) \frac{(y-r)}{y}  \,,
\label{total_width}
\end{equation}
for nonvanishing $m_1$. Again, the limit $r\to 0$ has to be taken in
the case of the decay $\tilde{\chi}^0_1\to b \bar{b}\nu_\tau$.

\subsection{Massive final state --- $\tilde{\chi}^0_1\to t \bar{b} \tau$}
\label{massive} 

\begin{figure}[ht] 
\begin{center}
\epsfxsize= 12 cm 
\leavevmode 
\epsfbox{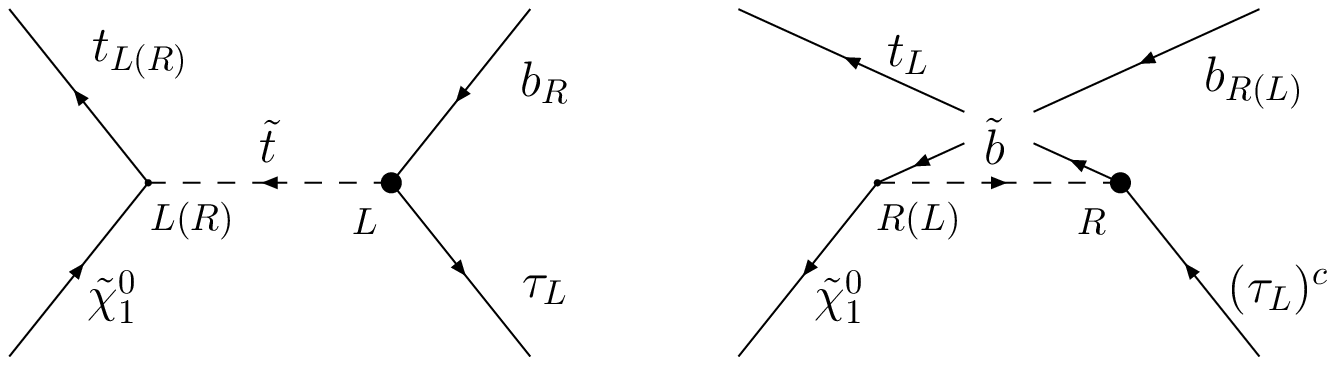}
\end{center} 
\caption{Diagrams
contributing to the decay $\tilde{\chi}^0_1\to t \bar{b} \tau$ through
the exchange of a virtual $\tilde{t}$- and $\tilde{b}$-squark.  The
thick vertex indicates the $R_p$-violating coupling
$\lambda_{333}^{\prime\ast}$.}\label{diagstopsbot}
\end{figure}

\begin{figure}[ht] 
\begin{center}
\epsfxsize= 3.4 cm 
\leavevmode 
\epsfbox{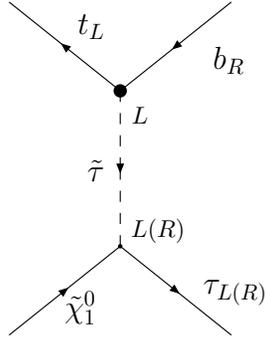}
\end{center} 
\caption{Diagram contributing to the decay
$\tilde{\chi}^0_1\to t \bar{b} \tau$ through exchange of a virtual
$\tilde{\tau}$-slepton.  The thick vertex indicates the
$R_p$-violating coupling
$\lambda_{333}^{\prime\ast}$.}\label{diagstau}
\end{figure}

This decay is mediated by the exchange of off-shell $\tilde{t}$- and
$\tilde{b}$-squarks, as shown in figure~\ref{diagstopsbot}, and an
off-shell $\tilde{\tau}$-slepton, as in figure~\ref{diagstau}. In
these figures, only the diagrams giving rise to matrix elements
proportional to $\lambda_{333}^{\prime\,\ast}$ are shown.  Notice
that, since the virtual $\tilde{\tau}$ exchanged in the diagram of
figure~\ref{diagstau} has both, left- and right-chiralities, there are
two independent contributions coming from this diagrams.  Once again,
we split the matrix elements for this decay as in
eq.~(\ref{ampl_decomp}).  The $Q$ terms relative to the diagrams in
figures~\ref{diagstopsbot} and~\ref{diagstau} are now \emph{six}
instead of \emph{five}:
\begin{eqnarray}
 Q_{S,RR}^x &=& 
 \left(\overline{u}_{\tau}(p_2) P_R\,v_{b}   (p_3)\right)
 \left(\overline{u}_{t}   (p_1) P_R\,u_{{\tilde{\chi}}^0_1}(q)\right)
\nonumber\\
 Q_{S,RL}^x &=& 
 \left(\overline{u}_{\tau}(p_2) P_R\,v_{b}   (p_3)\right)
 \left(\overline{u}_{t}   (p_1) P_L\,u_{{\tilde{\chi}}^0_1}(q)\right)
\nonumber\\
 Q_{S,RR}^z &=& 
 \left(\overline{u}_{\tau}(p_2) P_R\,v_{t}   (p_1)\right)
 \left(\overline{u}_{b}   (p_3) P_R\,u_{{\tilde{\chi}}^0_1}(q)\right)
\nonumber\\
 Q_{S,RL}^z &=& 
 \left(\overline{u}_{\tau}(p_2) P_R\,v_{t}   (p_1)\right)
 \left(\overline{u}_{b}   (p_3) P_L\,u_{{\tilde{\chi}}^0_1}(q)\right)
\nonumber\\
 Q_{S,RR}^y &=&  
 \left(\overline{u}_{t}   (p_1) P_R\,v_{b}   (p_3)\right)
 \left(\overline{u}_{\tau}(p_2) P_R\,u_{{\tilde{\chi}}^0_1}(q)\right)
\nonumber\\
 Q_{S,RL}^y &=& 
 \left(\overline{u}_{t}   (p_1) P_R\,v_{b}   (p_3)\right)
 \left(\overline{u}_{\tau}(p_2) P_L\,u_{{\tilde{\chi}}^0_1}(q)\right).
\label{q_diagr-mass} 
\end{eqnarray}
For convenience, the same symbols employed in eq.~(\ref{q_diagr}) are
used for these $Q$ terms, in spite of the fact that they are built out
of spinors of different particles. The squared products, summed over
all possible spin configurations obtained for the two sets of $Q$'s
(in eqs.~(\ref{q_diagr}) and~(\ref{q_diagr-mass})) differ only for the
value of the parameter $r = m_t^2/m_{\tilde{\chi}^0_1}$, which is
nonvanishing in the case of the decay $\tilde{\chi}^0_1\to t \bar{b}
\tau$.  The corresponding coefficients are now denoted by the symbols
$b^t_{S,s}$ and are:
{\setlength{\arraycolsep}{1.5pt}
\begin{equation}
\begin{array}{lllll}
b_{S,RR}^x  &      =           & +\sqrt{2}\,g
 \lambda_{333}^{\prime \ast}
 \left((L^{t_L})^{\ast}D_{LL}^{\tilde{t}}(x)
      +(L^{t_R})^{\ast}D_{RL}^{\tilde{t}}(x) \right)
            &      \to         & +\sqrt{2}\,g_Y 
 \lambda_{333}^{\prime \ast}\,\eta_1^{\ast}\,Y_{t_L}\,
 D_{LL}^{\tilde{t}}(x) 
\nonumber\\[1.01ex]     
b_{S,RL}^x  &      =           & -\sqrt{2}\,g
 \lambda_{333}^{\prime \ast}
 \left((R^{t_R})^{\ast}D_{RL}^{\tilde{t}}(x) 
      +(R^{t_L})^{\ast}D_{LL}^{\tilde{t}}(x) \right)
            &      \to         & -\sqrt{2}\,g_Y 
 \lambda_{333}^{\prime \ast}\,\eta_1^\ast\,Y_{t_R}\,
 D_{RL}^{\tilde{t}}(x)
\nonumber\\[1.01ex]
b_{S,RR}^z  &      =           & +\sqrt{2}\,g
 \lambda_{333}^{\prime \ast}
 \left((R^{b_R})(D_{RR}^{\tilde{b}}(z))^{\ast}
      +(R^{b_L})(D_{LR}^{\tilde{b}}(z))^{\ast} \right)
            &      \to         & +\sqrt{2}\,g_Y 
 \lambda_{333}^{\prime \ast}\,\eta_1\,Y_{b_R}
 (D_{RR}^{\tilde{b}}(z))^{\ast}
\nonumber\\[1.01ex]
b_{S,RL}^z  &      =           & -\sqrt{2}\,g
 \lambda_{333}^{\prime \ast}
 \left((L^{b_L})(D_{LR}^{\tilde{b}}(z))^{\ast}
      +(L^{b_R})(D_{RR}^{\tilde{b}}(z))^{\ast} \right)
            &      \to         & -\sqrt{2}\,g_Y 
 \lambda_{333}^{\prime \ast}\,\eta_1\,Y_{b_L}\,
 (D_{LR}^{\tilde{b}}(z))^{\ast} 
\nonumber\\[1.01ex]
b_{S,RR}^y  &      =           & -\sqrt{2}\,g
 \lambda_{333}^{\prime \ast}
 \left((L^{\tau_L})^{\ast}D_{LL}^{\tilde{\tau}}(y)
      +(L^{\tau_R})^{\ast}D_{RL}^{\tilde{\tau}}(y) \right)
            &      \to         & -\sqrt{2}\,g_Y 
 \lambda_{333}^{\prime \ast}\,\eta_1^{\ast}\,Y_{\tau_L}\,
 D_{LL}^{\tilde{\tau}}(y) 
\nonumber\\[1.01ex]   
b_{S,RL}^y  &      =           & +\sqrt{2}\,g
 \lambda_{333}^{\prime \ast}
 \left((R^{\tau_R})^{\ast}D_{RL}^{\tilde{\tau}}(y)
      +(R^{\tau_L})^{\ast}D_{LL}^{\tilde{\tau}}(y) \right)
            &      \to         & +\sqrt{2}\,g_Y 
 \lambda_{333}^{\prime \ast}\,\eta_1^\ast\,Y_{\tau_R}\,
 D_{RL}^{\tilde{\tau}}(y)                  \,.
\end{array}
\label{coeff_diagr-mass}
\end{equation}}
Notice that no correspondent of the crossed diagrams considered in 
the massless final state case exist here. 

\subsection{Massive final state ---
  $\tilde{\chi}^0_1\to \bar{t} {b} \bar{\tau}$}
\label{massive_conj}

This decay mode is obtained $CP$ conjugating the previous one
$\tilde{\chi}^0_1\to t \bar{b} \tau$, and it has therefore the same
width as the previous one. For completeness, however, we give also the
matrix elements for this decay. Splitting them again into $Q$ terms
and coefficients, we find:
\begin{eqnarray}
 Q_{S,LL}^x &=& 
 \left(\overline{u}_{\tau}(p_2) P_L v_{b}   (p_3)\right)
 \left(\overline{u}_{t}   (p_1) P_L\,u_{{\tilde{\chi}}^0_1}(q)\right)
\nonumber\\[1.01ex]
 Q_{S,LR}^x &=& 
 \left(\overline{u}_{\tau}(p_2) P_L v_{b}   (p_3)\right)
 \left(\overline{u}_{t}   (p_1) P_R\,u_{{\tilde{\chi}}^0_1}(q)\right)
\nonumber\\[1.01ex]
 Q_{S,LL}^z &=& 
 \left(\overline{u}_{\tau}(p_2) P_L v_{t}   (p_1)\right)
 \left(\overline{u}_{b}   (p_3) P_L\,u_{{\tilde{\chi}}^0_1}(q)\right)
\nonumber\\[1.01ex]
 Q_{S,LR}^z &=& 
 \left(\overline{u}_{\tau}(p_2) P_L v_{t}   (p_1)\right)
 \left(\overline{u}_{b}   (p_3) P_R\,u_{{\tilde{\chi}}^0_1}(q)\right)
\nonumber\\[1.01ex]
 Q_{S,LL}^y &=&  
 \left(\overline{u}_{t}   (p_1) P_L v_{b}   (p_3)\right)
 \left(\overline{u}_{\tau}(p_2) P_L\,u_{{\tilde{\chi}}^0_1}(q)\right)
\nonumber\\[1.01ex]
 Q_{S,LR}^y &=& 
 \left(\overline{u}_{t}   (p_1) P_L v_{b}   (p_3)\right)
 \left(\overline{u}_{\tau}(p_2) P_R\,u_{{\tilde{\chi}}^0_1}(q)\right)\,.
\label{q_cpdiagr-mass} 
\end{eqnarray}
Notice that these differ form the $Q$ terms in
eq.~(\ref{q_diagr-mass}) for an exchange $R \leftrightarrow L$.  The
corresponding coefficients,
{\setlength{\arraycolsep}{1.4pt} 
\begin{equation}
\begin{array}{lllll}
c_{S,LL}^x  &\!    =            & +\sqrt{2}\,g
 \lambda_{333}^{\prime}
 \left((L^{t_L}) (D_{LL}^{\tilde{t}}(x))^{\ast}
      +(L^{t_R}) (D_{RL}^{\tilde{t}}(x))^{\ast} \right)
            &\!    \to          & +\sqrt{2}\,g_Y 
 \lambda_{333}^{\prime}\,\eta_1\, Y_{t_L}
 \left(D_{LL}^{\tilde{t}}(x)\right)^{\ast} 
\nonumber\\[1.01ex]     
c_{S,LR}^x  &\!    =            & -\sqrt{2}\,g
 \lambda_{333}^{\prime}
 \left((R^{t_R}) (D_{RL}^{\tilde{t}}(x))^{\ast}
     +(R^{t_L}) (D_{LL}^{\tilde{t}}(x))^{\ast} \right)
            &\!    \to          & -\sqrt{2}\,g_Y 
 \lambda_{333}^{\prime}\,\eta_1\, Y_{t_R} 
 \left(D_{RL}^{\tilde{t}}(x)\right)^{\ast} 
\nonumber\\[1.01ex]          
c_{S,LL}^z  &\!    =            & +\sqrt{2}\,g
 \lambda_{333}^{\prime}
 \left((R^{b_R})^{\!\ast}D_{RR}^{\tilde{b}}(z)
      +(R^{b_L})^{\!\ast}D_{LR}^{\tilde{b}}(z) \right)
            &\!    \to          & +\sqrt{2}\,g_Y 
 \lambda_{333}^{\prime}\,\eta_1^\ast\, Y_{b_R} 
 \left(D_{RR}^{\tilde{b}}(z)\right)
\nonumber\\[1.01ex]          
c_{S,LR}^z  &\!    =            & -\sqrt{2}\,g
 \lambda_{333}^{\prime}
 \left((L^{b_L})^{\!\ast}D_{LR}^{\tilde{b}}(z)
      +(L^{b_R})^{\!\ast}D_{RR}^{\tilde{b}}(z) \right)
            &\!    \to          & -\sqrt{2}\,g_Y 
 \lambda_{333}^{\prime}\,\eta_1^{\ast}\, Y_{b_L} 
 \left(D_{LR}^{\tilde{b}}(z)\right)
\nonumber\\[1.01ex]          
c_{S,LL}^y  &\!    =            & -\sqrt{2}\,g
 \lambda_{333}^{\prime}
 \left((L^{\tau_L})(D_{LL}^{\tilde{\tau}}(y))^{\ast}
      +(L^{\tau_R})(D_{RL}^{\tilde{\tau}}(y))^{\ast} \right)
            &\!    \to          & -\sqrt{2}\,g_Y 
 \lambda_{333}^{\prime}\,\eta_1\, Y_{\tau_L} 
 \left(D_{LL}^{\tilde{\tau}}(y)\right)^{\ast}
\nonumber\\[1.01ex]             
c_{S,LR}^y  &\!    =            & +\sqrt{2}\,g
 \lambda_{333}^{\prime}
 \left((R^{\tau_R})(D_{RL}^{\tilde{\tau}}(y))^{\ast}
      +(R^{\tau_L})(D_{LL}^{\tilde{\tau}}(y))^{\ast} \right)
            &\!    \to          & +\sqrt{2}\,g_Y 
 \lambda_{333}^{\prime}\,\eta_1\, Y_{\tau_R} 
 \left(D_{RL}^{\tilde{\tau}}(y)\right)^{\ast} ,
\end{array}
\label{coeff_cpdiagr-mass}
\end{equation}}
are, as expected:
\begin{equation}
 c_{S,LL}^t = \left(b_{S,RR}^{t}\right)^{\ast},
\qquad
 c_{S,LR}^t = \left(b_{S,RL}^{t}\right)^{\ast},
\end{equation}
where $t = x,y,z$.

\section{Additional large couplings of type  $\lambda^\prime$ 
 and $\lambda$} 
\label{othercouplings}

As argued in section~\ref{limits}, there are other couplings of type
$\lambda$ and $\lambda^\prime$, besides $\lambda^\prime_{333}$, that
can be large.  They originate a variety of final states with possibly
interesting experimental signatures. We start discussing some final
states due to couplings of type $\lambda^\prime$, in particular those
with two indices of third generation and one of second
generation. Generalization to other cases are obvious.

\begin{figure}[ht] 
\begin{center}
\epsfxsize= 12 cm 
\leavevmode 
\epsfbox{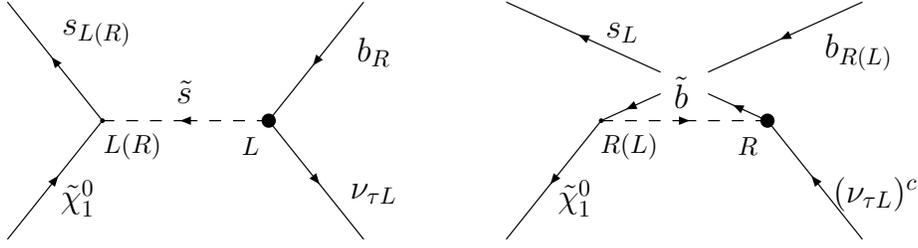}
\end{center} 
\caption{Diagrams
contributing to the decay $\tilde{\chi}^0_1\to s \bar{b} \nu_\tau$
through the exchange of a virtual $\tilde{s}$- and $\tilde{b}$-squark.
The thick vertex indicates the $R_p$-violating coupling
$\lambda_{323}^{\prime\ast}$.}\label{diagsstrsbot}
\end{figure}

\begin{figure}[ht] 
\begin{center}
\epsfxsize= 3.4 cm 
\leavevmode 
\epsfbox{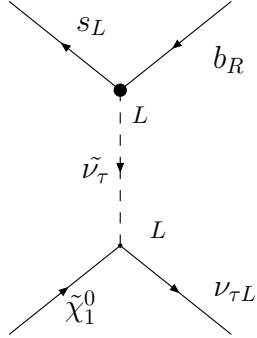}
\end{center} 
\caption{Diagram
contributing to the decay $\tilde{\chi}^0_1\to s \bar{b} \nu_\tau$
through exchange of a virtual $\tilde{\nu_\tau}$-slepton.  The thick
vertex indicates the $R_p$-violating coupling
$\lambda_{323}^{\prime\ast}$.}\label{diagsnutau}
\end{figure}

A nonvanishing coupling $\lambda^\prime_{323}$ gives rise to final
states with only light particles. There are two decays into charmed
final states: $\tilde{\chi}^0_1 \to c \bar{b} \tau$ and
$\tilde{\chi}^0_1 \to \bar{c} b \bar{\tau}$. The diagrams relative to
the decay $\tilde{\chi}^0_1 \to c \bar{b} \tau$ can be obtained from
those in figures~\ref{diagstopsbot} and~\ref{diagstau}, substituting
everywhere $t\to c$. The matrix elements are obtained from those in
eqs.~(\ref{q_diagr-mass}) and~(\ref{coeff_diagr-mass}), and
in~(\ref{q_cpdiagr-mass}) and~(\ref{coeff_cpdiagr-mass}),
respectively. In the evaluation of the matrix elements squared,
however, the limit $r\to 0$ has to be taken in
appendix~\ref{squaredoperators}.  Decay modes with $s$-quark in the
final state, i.e.\ $s \bar{b} \nu_\tau$ and $\bar{s} b \nu_\tau$ are
also possible. The Feynmman diagrams for $\tilde{\chi}^0_1 \to s
\bar{b} \nu_\tau$ due to the coupling $\lambda^{\prime\ast}_{323}$ are
shown explicitly in figures~\ref{diagsstrsbot} and~\ref{diagsnutau}.
(Those due to the coupling $\lambda^{\prime\ast}_{332}$ are obtained
from the diagrams in figures~\ref{diagsstrsbot} and~\ref{diagsnutau}
interchanging $s$ with $b$ everywhere.)  The matrix elements for the
decay mode $\tilde{\chi}^0_1 \to s \bar{b} \nu_\tau$ can be obtained
with the replacement $b \to s$ in the first and second line of
eq.~(\ref{coeff_diagr}) and by taking the momentum $p_1$ in
eq.~(\ref{q_diagr}) to be the momentum of the $s$-quark, that is to
say, substituting $u_b(p_1)$ with $u_s(p_1)$.  Those for the decay
mode $\tilde{\chi}^0_1 \to \bar{s} b \nu_\tau$, can be obtained by
replacing $b \to s$ in the third and forth line of
eq.~(\ref{coeff_crossdiagr}) and by taking the momentum $p_3$ in
eq.~(\ref{q_crossdiagr}) to be the momentum of the $s$-quark, i.e.\
substituting $u_b(p_3)$ with $u_s(p_3)$.

As in the case of $\lambda^\prime_{333}$, the coupling
$\lambda^\prime_{233}$ induces also decays of the lightest neutralino
into a roughly massless final state, $\bar{b} b \nu_\mu$, and into
massive ones: $\bar{b} t \mu$ with its conjugate state $b
\bar{t}\bar{\mu}$.  All formulae derived in
section~\ref{lambdaprime333} apply also to this case, with the obvious
changes: $\nu_\tau \to \nu_\mu$, $\tau \to \mu$, $\tilde{\tau} \to
\tilde{\mu}$ and $\lambda^\prime_{333} \to \lambda^\prime_{233}$.

\looseness=-1 Similarly, a nonvanishing coupling
$\lambda^\prime_{332}$ gives also rise to decay modes with massive
particles in the final state: $t\bar{s} \tau$ and $\bar{t} s
\bar{\tau}$. The treatment of these decays follows that for $t \bar{b}
\tau$ and $\bar{t} b \bar{\tau}$ in sections~\ref{massive}
and~\ref{massive_conj}. (All needed formulae are obtained from those
in these sections with the substitution $b\to s$, $\tilde{b} \to
\tilde{s}$, and $\lambda_{333}^\prime \to \lambda_{332}^\prime$.) The
massless final states induced by this coupling are $\bar{b} s
\nu_\tau$ and $b \bar{s} \nu_\tau $, already discussed in the case of
the coupling $\lambda_{323}^\prime$.  However, $\lambda^\prime_{323}$
naturally induces a left-handed $s$-field, whereas
$\lambda^\prime_{332}$ induces a right-handed one. The matrix elements
for the decay $\tilde{\chi}^0_1\to b \bar{s} \nu_\tau $ can be gleaned
from the $Q$ terms and coefficients in eqs.~(\ref{q_diagr})
and~(\ref{coeff_diagr}) as follows: substitute $b \to s$ in the third
and fourth line of eq.~(\ref{coeff_diagr}) and take the momentum $p_3$
in eq.~(\ref{q_diagr}) to be the momentum of the $s$-quark, that is to
say, substitute $v_b(p_3)$ with $v_s(p_3)$.  The $Q$ terms and
coefficients in eqs.~(\ref{q_crossdiagr}), in turn, can be used to
obtain the matrix elements for the decay $\tilde{\chi}^0_1 \to s
\bar{b} \nu_\tau $. The replacement recipe is now as follows:
substitute $b \to s$ in the first and second line of
eq.~(\ref{coeff_crossdiagr}) and take the momentum $p_1$ in
eq.~(\ref{q_crossdiagr}) to be the momentum of the $s$-quark, i.e.\
substitute $u_b(p_1)$ with $u_s(p_1)$.

Notice that, among the abovementioned couplings, only
$\lambda^\prime_{323}$ and $\lambda^\prime_{332}$ remain unconstrained
by the requirement that the loop contributions to neutrino masses are
not too large, if the left-right mixing terms in the sfermion sector
do not play a role in the suppression of these loops.


The typical signatures induced by the coupling $\lambda^\prime_{333}$
that may be expected at the LHC were already listed in
section~\ref{intro}. In a similar way, the couplings
$\lambda^\prime_{332}$ and $\lambda^\prime_{323}$ give rise to the
$R_p$-conserving final states
\begin{eqnarray}
pp  \to t \bar{b} \tilde{\mu} X 
   & \to & t \bar{b} \bar{t} b   X \,, 
\nonumber\\
pp  \to t \bar{s} \tilde{\tau}X 
   & \to & t \bar{s} \,\bar{t} s X \,, 
\nonumber\\
pp  \to c \bar{b} \tilde{\tau} X
   & \to & c \bar{b} \,\bar{c} b  X \,,
\label{bosonsignal} 
\end{eqnarray}
which may also be induced by flavor-conserving and flavor-violating
decays of charged-Higgs bosons, or, in the last case, by
flavor-violating decays of a pair of $W$ bosons. The couplings
$\lambda^\prime_{233}$, $\lambda^\prime_{332}$, and 
$\lambda^\prime_{323}$ also originate $R_p$-violating final states 
such as 
\begin{eqnarray}
pp  \to t \bar{b} \tilde{\mu} X 
   & \to &  (2t) \,(2\bar{b})\, (2\mu)  X\,, 
\nonumber\\
pp  \to t \bar{s} \tilde{\tau}X 
   & \to &  (2t) \,(2\bar{s})\, (2\tau) X\,,
\nonumber\\
pp  \to c \bar{b} \tilde{\tau} X
   & \to &  (2c) \,(2\bar{b})\, (2\tau) X\,, \qquad 
\label{addrpsignals} 
\end{eqnarray}
giving rise to pairs of like-sign dileptons. As already mentioned in
section~\ref{intro}, that of like-sign dileptons constitute a
distinctive signature of these production and decay mechanisms when
the $t$-quarks involved in these final states decay completely into
hadrons.  The $R_p$-violating states $(2t)\,(2\bar{b})\, (2\mu)$,
$(2c)\,(2\bar{b})\, (2\tau)$, $(2t)\,(2\bar{s})\, (2\tau)$, and their
conjugated ones, are the typical final states obtained from
pair-produced lightest neutralinos, (at the LHC or $e^+e^-$
colliders), when couplings $\lambda^\prime_{ijk}$ with more than two
third-generation indices are nonvanishing.

Couplings of type $\lambda$ are antisymmetric in the first two
indices. Therefore, there is only one such coupling with two
third-generation indices: $\lambda_{233}$. The neutralino decays
induced by this coupling give rise to the final states: $\tau
\bar{\tau} \nu_\mu$, and $\mu \nu_\tau \bar{\tau}$, together with
$\bar{\mu} \nu_\tau {\tau}$. (Once again, no distinction is made
between $\nu_\tau$ and $\bar{\nu}_\tau$.) The calculations of the
corresponding widths match those in section~\ref{lambdaprime333}, when
the following changes are made: $b \to \tau$, $\tilde{b} \to
\tilde{\tau}$, $\nu_\tau\to \nu_\mu$, and $\lambda^\prime_{333} \to
\lambda_{233}$, in the case of the first decay mode, $\tau \bar{\tau}
\nu_\mu$; and $t\to\nu_\tau$, $\tau \to \mu$, $b \to \tau$, and again
$\lambda^\prime_{333} \to \lambda_{233}$, in the case of the remaining
two decay modes $\mu \nu_\tau \bar{\tau}$ and $\bar{\mu} \nu_\tau
{\tau}$. Typical final states that can be expected at the LHC are:
\begin{eqnarray}
pp  \to t \bar{b} \tilde{\tau} X 
   & \to &  t \bar{b}(2\tau) \bar{\tau} \nu_\nu X\,, 
\nonumber\\
   & \to &  t \bar{b}(2\tau) \bar{\mu} \nu_\tau X\,;
\label{rpsignalslambda} 
\end{eqnarray}
whereas signatures such as $(2\tau) (2 \bar{\tau})$ and missing
energy, or the typical like-sign dilepton signatures
$(2\tau)(2\bar{\mu})$ (and $(2\bar{\tau})(2{\mu})$) plus missing
energy are obtained from pair-produced lightest neutralinos.

The discussion can be generalized to other couplings of type $\lambda$
with more than one index different from $3$.

\section{Numerical results} 
\label{results}

We are now in a position to discuss the relative size of widths for
the final states originated by the decay of the lightest neutralino,
when some of the $\lambda^\prime$ and $\lambda$ couplings are
considerably larger than the others.  We concentrate first on the
somewhat ideal case in which only one $R_p$-violating coupling is
present, $\lambda_{333}^\prime$, and $\tilde{\chi}^0_1$ can only decay
into one of the three final states $b \bar{b} \nu_{\tau}$, $t \bar{b}
\tau$, and $\bar{t} {b} \bar{\tau}$.  We shall discuss later other
decay modes originated by other couplings of type $\lambda^\prime$.
Finally we shall show results for a framework in which
$\lambda_{333}^\prime$ and $\lambda_{233}$ are simultaneously
nonvanishing, and dominant among all $R_p$-violating couplings.

In general, we do not assume the relation $m_{\tilde{W}} \simeq 2
m_{\tilde{B}}$ among gaugino masses, which is typical of an mSUGRA
scenario, with gaugino mass unification at a high scale. Indeed,
gaugino mass unification is not a necessary ingredient of this
scenario, only a customary one. Moreover, models with Wino states
lighter than the Bino have recently emerged. The relation
$m_{\tilde{B}} = k m_{\tilde{W}}$, with $k> 1$, is, for example,
predicted in models in which the breaking of supersymmetry is
transmitted to the visible sector through anomaly meditation
(see~\cite{Gherghetta:1999sw} and references therein). In this case,
it is $k \simeq 3$. A more complicated relation between
$m_{\tilde{B}}$ and $m_{\tilde{W}}$ is also possible in some grand
unification models~\cite{NEWUNIF} in which the usual gaugino mass
unification in replaced by a more complicated relation among the
gaugino masses~\cite{NEWGAUGINOREL}.  In this case, $m_{\tilde{W}}$
and $m_{\tilde{B}}$, together with the gluino mass, $m_{\tilde{g}}$,
satisfy a linear constraint, $m_{\tilde{B}}= k m_{\tilde{W}} + h
m_{\tilde{g}}$, with a wide range of values possible for $k$.

In the following, only the two specific values $k=1/2$ and $k=2$ will
be discussed.

\subsection{Only $\lambda_{333}^\prime\ne 0$ }
\label{reslprime333}

For a relatively light lightest neutralino, i.e.\ when
${\tilde\chi}_1^0$ is below the $t$-quark threshold, the only possible
decay mode is $b \bar{b} \nu_{\tau}$.  Above the $t$-quark threshold,
we should distinguish the case in which $\tilde{\chi}^0_1$ is mainly a
gaugino with a small contamination from the two
Higgsinos,
from the case in which $\tilde{\chi}^0_1$ has
a substantial Higgsino component. In the former, among the three
states in which ${\tilde\chi}_1^0$ can decay, $b \bar{b} \nu_{\tau}$,
$t \bar{b} \tau$, and $\bar{t} b \bar{\tau}$, the state $b \bar{b}
\nu_{\tau}$ has the largest width and therefore the largest branching
ratio, when the values of the sfermion masses virtually exchanged in
the three decays are not too dissimilar.  The branching ratio of the
two combined massive modes, however, is, in a wide range of parameter
space, of the same order of magnitude of that for the $b \bar{b}
\nu_{\tau}$ mode. This statement is valid irrespective of the value of
$\tan \beta$.  We observe that the widths for all decay modes are
larger if the main gaugino component of ${\tilde\chi}_1^0$ is of Wino
type instead than of Bino type.

The two massive modes may become dominant if the gaugino-Higgsino
admixture in the lightest neutralino state is substantial and $\tan
\beta$ is not too large. This is due to the large $t$-quark Yukawa
coupling. When $\tan \beta$ increases and the $b$-quark Yukawa
coupling becomes comparable to that of the $t$-quark, the decay mode
$b \bar{b} \nu_{\tau}$ takes over and dominates over the $t \bar{b}
\tau$ one.

\begin{figure}[t] 
\begin{center}
\epsfxsize= 9 cm 
\leavevmode 
\epsfbox{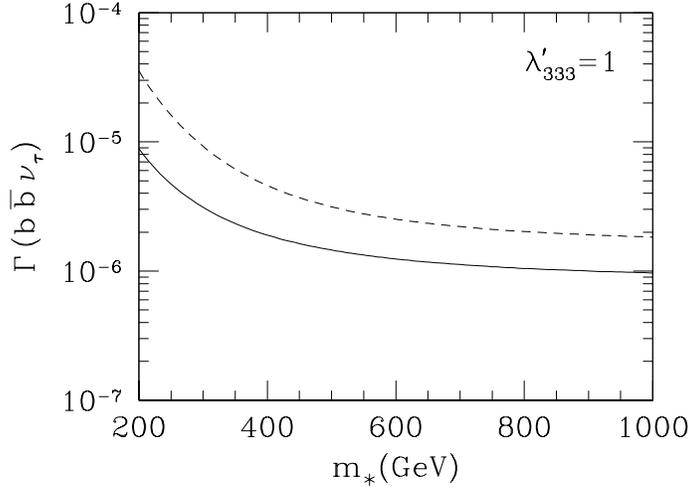}
\end{center} 
\caption[f1]{$\Gamma(\tilde{\chi}^0_1 \to b \bar{b}
\nu_{\tau})$ (in GeV) versus
$m_\ast = m_{\tilde{Q}}=m_{\tilde{U^c}}=m_{\tilde{D^c}}$ for the
decay of an almost pure gaugino ${\tilde\chi}_1^0$ for $\mu=
500\,$GeV, $\tan \beta = 3$, $m_{\tilde{L}}=m_{\tilde{E^c}}=400\,$GeV,
and gaugino masses $m_{\tilde{B}} = (1/2)m_{\tilde{W}} = 100\,$GeV
(solid lines), $m_{\tilde{W}} = (1/2)m_{\tilde{B}} = 100\,$GeV
(dashed lines).  The trilinear soft terms were chosen in such a way to
cancel the left-right mixing terms in all sfermion mass matrices.}
\label{bvsw-lchi}
\end{figure}

We show in figure~\ref{bvsw-lchi} the width $\Gamma(\tilde{\chi}^0_1
\to b \bar{b} \nu_{\tau})$, in the case of a light ${\tilde\chi}_1^0$
that is mainly a Bino (solid line), or mainly a Wino (dashed line). In
the lower curve, the $\tilde{B}$ mass is fixed to $100\,$GeV, the
$\tilde{W}$ mass is assumed to be twice as large as the $\tilde{B}$
mass, whereas the $\mu$ parameter is given the larger value of
$500\,$GeV.  The width is shown as a function of the doublet squark
mass, $m_{\tilde{Q}}$ assumed to be equal to both, the stop and
sbottom singlet masses, $m_{\tilde{U^c}}$ and $m_{\tilde{D^c}}$.  The
slepton doublet and singlet masses $m_{\tilde{L}}=m_{\tilde{E^c}}$ are
fixed to $400\,$GeV.  The soft trilinear terms for squarks and
sleptons as well as $\tan \beta$ are chosen in such a way to obtain
vanishing left-right entries in the corresponding sfermion mass
matrices.  Specifically, $\tan \beta =3$ was used for this
figure. This choice, however, does not affect the results shown here:
the width $\Gamma(\tilde{\chi}^0_1 \to b \bar{b} \nu_{\tau})$ is
practically independent of the value of $\tan \beta$, in the
approximation of $\tilde{\chi}^0_1$ as a pure gaugino and when the
left-right mixing terms among sfermions are vanishing. In this case,
the only dependence on $\tan \beta$ of the widths for the lightest
neutralino decays is, in principle, due to the $D$-term contributions
to the sfermion mass eigenvalues. For left- and right-entries in the
sfermion mass matrices larger than $200\,$GeV, this dependence is
completely negligible.  Thus, for simplicity, in this figure and in
the following ones, the $D$-term contributions to the sfermion mass
matrices were ignored.  Furthermore, entries in the sfermion mass
matrices that are explicitly $R_p$-violating, are proportional to the
bilinear couplings $k_i$ and to the \emph{vev}'s $v_i$ of the fields
$\tilde{L}_i$. They were dropped in this analysis, since completely
negligible.

The upper curve in this plot (dashed line) shows the width
\smash{$\Gamma(\tilde{\chi}^0_1 \to b \bar{b} \nu_{\tau})$} in the case in
which $\tilde{\chi}^0_1 $ is mainly a Wino. This approximation was
achieved simply inverting the ratio of $m_{\tilde{B}}$ and
$m_{\tilde{W}}$, i.e.\ taking $m_{\tilde{B}} = 2 m_{\tilde{W}}$.  All
other massive parameters have the same values as those chosen in the
approximation in which the lightest neutralino is mainly a Bino.
Although one diagram less contributes now to \smash{$\Gamma(\tilde{\chi}^0_1
\to b \bar{b}\nu_{\tau})$} (the right-handed sbottom squarks does not
couple to the Wino), the larger value of the couplings
fermion-sfermion-Wino accounts for the larger width obtained in this
case.

It should be mentioned here that some of the trilinear terms used in
this figure (as well as in some of the following figures) to cancel
the left-right sfermion mixing terms, may seem, at times, a little too
large. However, as it was argued in ref.~\cite{Borzumati:1999sp},
there are ways to accommodate large trilinear terms in supersymmetric
models, provided they do not give rise to negative sfermion mass
squared eigenvalues. One of the main constraints to their size
remains, the requirement that the radiative contribution to fermion
masses induced by such large terms does not exceed the experimentally
observed ones.
\begin{figure}[h] 
\begin{center}
\epsfxsize= 9 cm 
\leavevmode 
\epsfbox{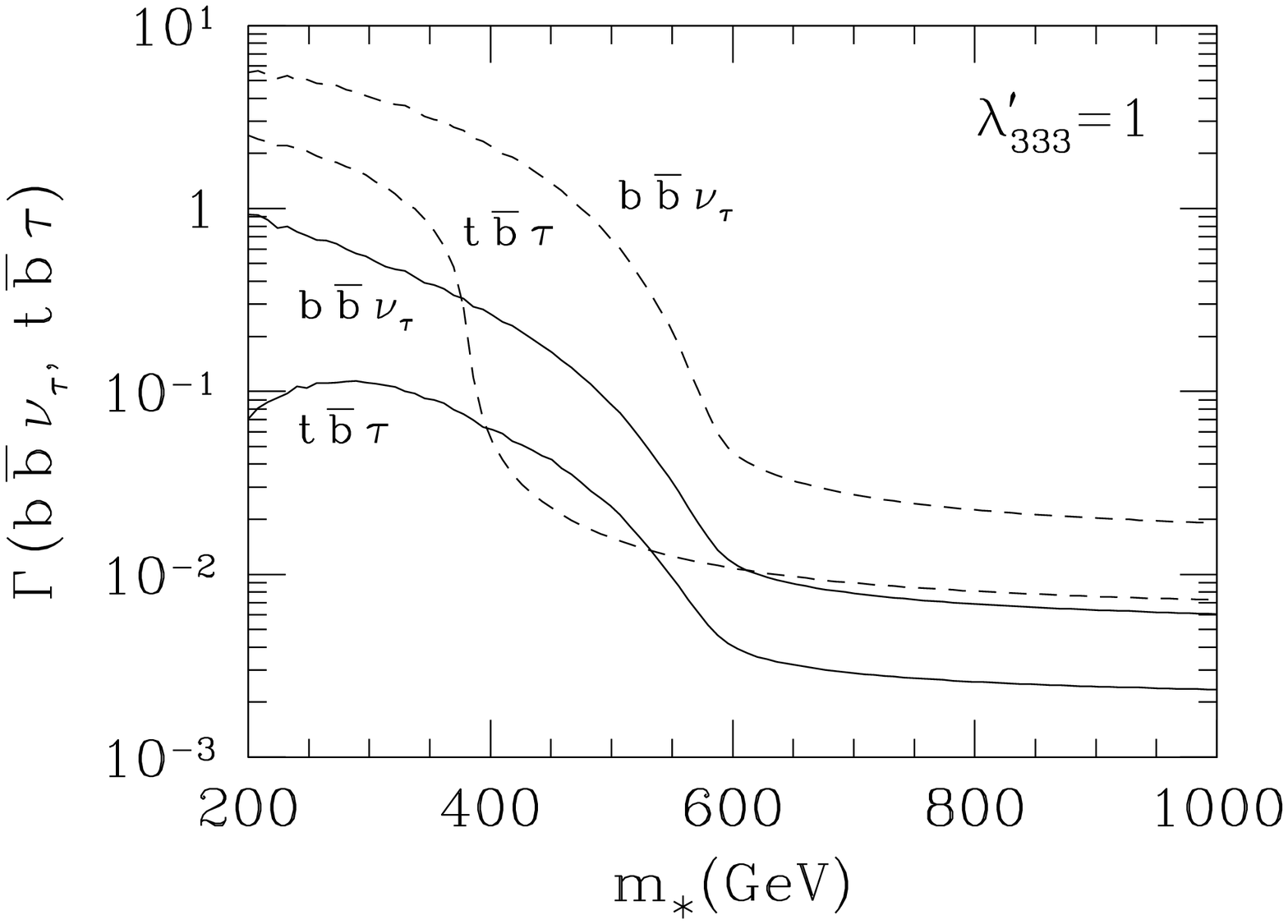} 
\vspace*{0.01truecm}
\epsfxsize= 9 cm 
\epsfbox{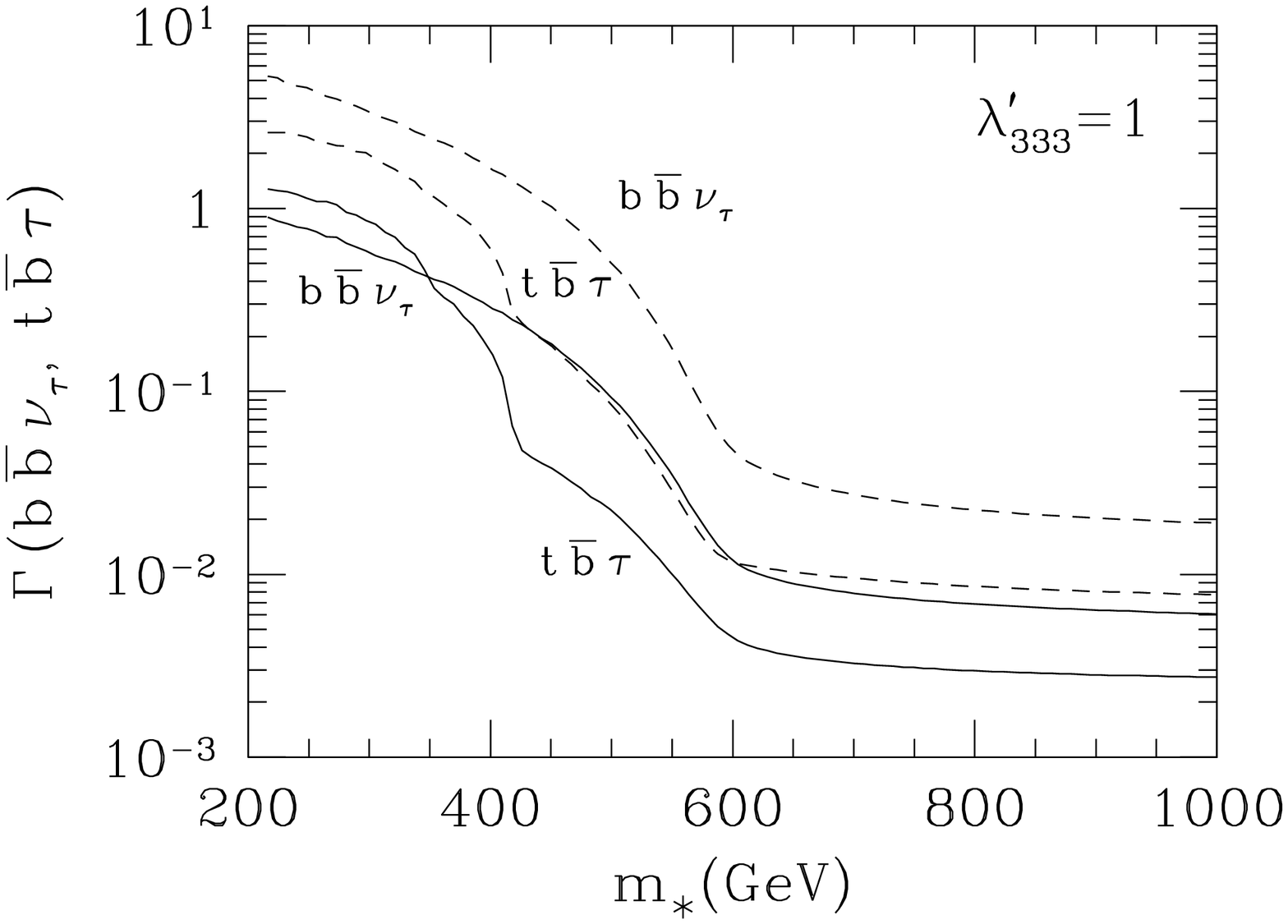} 
\end{center} 
\caption{$\Gamma(\tilde{\chi}^0_1 \to  b \bar{b}
\nu_{\tau})$ and $\Gamma(\tilde{\chi}^0_1 \to  t \bar{b} {\tau})$
(in\,GeV) (in\,GeV) versus
$m_\ast = m_{\tilde{Q}} = m_{\tilde{U^c}} = m_{\tilde{D^c}}$,
for $\mu =  1500\,$GeV, $\tan \beta =  3$,
$m_{\tilde{L}} = m_{\tilde{E^c}}  = 600\,$GeV. The gaugino masses
are $m_{\tilde{B}} = (1/2)m_{\tilde{W}} = 600\,$GeV (solid lines),
$m_{\tilde{W}} = (1/2)m_{\tilde{B}} = 600\,$GeV (dashed lines).
The trilinear soft terms are chosen in such a way to give vanishing
left-right entries in all sfermion mass matrices in the upper frame
and moderate left-right entries the lower frame.}
\label{bvsw-hchi}
\end{figure}

In figure~\ref{bvsw-hchi} the almost complete dominance of the widths
$\Gamma(\tilde{\chi}^0_1 \to b \bar{b} \nu_{\tau})$ and
$\Gamma(\tilde{\chi}^0_1 \to t \bar{b} {\tau})$, obtained for a
\smash{$\tilde{\chi}^0_1$} that is mainly a Wino, over those obtained when
\smash{$\tilde{\chi}^0_1$} is mainly a Bino, is explicitly shown in the case
of a heavy lightest neutralino, i.e.\ with mass $\simeq
600\,$GeV. Indeed, the lightest of the two gauginos, the Bino in the
solid lines ($m_{\tilde{W}}=2 m_{\tilde{B}}$), or the Wino in the
dashed lines ($m_{\tilde{B}}= 2 m_{\tilde{W}}$), are fixed at
$600\,$GeV, whereas the $\mu$ parameter has the large value
$\mu=1500\,$GeV. The values $m_{\tilde{L}}=m_{\tilde{E^c}}=600\,$GeV
are used for the slepton masses. No left-right mixing in all sfermion
mass matrices is assumed in the upper frame of this figure. This is
achieved fixing the value of $\tan \beta$ to $3$ and choosing
consequently the trilinear $A$ couplings. As already mentioned, for a
lightest neutralino that is mainly a gaugino, there is practically no
$\tan \beta$ dependence in the widths \smash{$\Gamma(\tilde{\chi}^0_1 \to b
\bar{b} \nu_{\tau})$}, \smash{$\Gamma(\tilde{\chi}^0_1 \to t \bar{b}
\tau)$}, and \smash{$\Gamma(\tilde{\chi}^0_1 \to \bar{t} b
\bar{\tau})$}, in the absence of left-right mixing terms among
sfermions. In both approximations, that of a mainly Bino and that of a
mainly Wino for the lightest neutralino, for similar $\tilde{t}$ and
$\tilde{b}$ eigenvalues, the decay mode \smash{$\tilde{\chi}^0_1 \to b
\bar{b} \nu_{\tau}$} dominates over the other two, which are penalized
by a large phase-space suppression and by the fact that they are here
considerated separately: being $b \bar{b}\nu_{\tau}$ a self-conjugated
state, the decay \smash{$\tilde{\chi}^0_1 \to b \bar{b} \nu_{\tau}$}
collects actually contributions to \smash{$\tilde{\chi}^0_1 \to b
\bar{b} \nu_{\tau}$} and \smash{$\tilde{\chi}^0_1 \to \bar{b} b
\bar{\nu}_{\tau}$}.  The upper curves in the two sets of solid and
dashed lines, show \smash{$\Gamma(\tilde{\chi}^0_1 \to b \bar{b}
\nu_{\tau})$} as a function of
\smash{$m_{\tilde{Q}}=m_{\tilde{U^c}}=m_{\tilde{D^c}}$}, and
\smash{$\Gamma(\tilde{\chi}^0_1 \to t \bar{b} \tau)$} is shown by the
two lower curves. In the dashed curve corresponding to
\smash{$\Gamma(\tilde{\chi}^0_1 \to t \bar{b} \tau)$}, which is
determined by the $\tilde{t}$- and $\tilde{\tau}$-exchange diagrams,
(no right-handed $\tilde{b}$ exchange is possible in this case, since
the lightest neutralino is mainly a Wino) the resonant region in which
the $\tilde{t}$-squark is produced on shell is clearly visible.
Between $400 \le m_{\tilde{Q}} \le 600\,$GeV, the width drops roughly
as $1/(m_{\tilde{t}}^2)^2$, whereas the almost flat behavior after
$600\,$GeV is due to the dominance of the $\tilde{\tau}$-exchange
diagram.  The solid line representing \smash{$\Gamma(\tilde{\chi}^0_1
\to t \bar{b} \tau)$}, and corresponding to a mainly Bino lightest
neutralino, is determined by all the three diagrams in
figure~\ref{diagstau}. In the resonant region in which
$\tilde{b}$-squarks are produced on shell, the diagram with exchange
of a right-handed $\tilde{b}$-squark is actually the dominant one. For
decreasing values of $m_{\tilde{Q}}$, in the region $m_{\tilde{Q}} <
600\,$GeV, the increase of the width, due to the dependence on
\smash{$(m_{\tilde{\chi}^0_1}^2 -m_{\tilde{b}}^2)^{-2}$} is damped at the
lower end of $m_{\tilde{Q}}$ by the severe phase-space suppression in
the branching ratio of the $2$-body decay \smash{$\tilde{b} \to t \tau$}.


In the lower frame of figure~\ref{bvsw-hchi}, a moderate splitting
among the two $\tilde{t}$ and $\tilde{b}$ eigenvalues is allowed,
i.e.\ $A_t -\mu /\tan \beta=150\,$GeV, and $(A_b -\mu \tan \beta)
m_b\sim (100)^2\,$GeV$^2$. Because the left-right mixing terms in the
$\tilde{b}$ mass matrix square are nonvanishing, the
$\tilde{b}$-mediated diagram contributes now to the width
$\Gamma(\tilde{\chi}^0_1 \to t \bar{b} \tau)$ also when the lightest
neutralino is mainly a Wino. This contribution explains the difference
in shape of the width $\Gamma(\tilde{\chi}^0_1 \to t \bar{b} \tau)$ in
the region $400 \le m_{\tilde{Q}} \le 600\,$GeV, with respect to that
obtained in the absence of left-right mixing terms. Similarly, the
smaller values of the mass of the lightest $\tilde{t}$- and
$\tilde{b}$-squarks explain the large enhancement of the width
$\Gamma(\tilde{\chi}^0_1 \to t \bar{b} \tau)$ obtained for
$m_{\tilde{Q}} \le 400\,$GeV when $\tilde{\chi}^0_1$ is mainly a
Bino. Larger values of $\tan\beta$, i.e.\ larger values of left-right
mixing terms in the $\tilde{b}$ mass matrix, would further enhance
$\Gamma(\tilde{\chi}^0_1\to b \bar{b} \nu_{\tau})$ with respect to
$\Gamma(\tilde{\chi}^0_1\to t \bar{b} {\tau})$.

\begin{figure}[ht] 
\begin{center}
\epsfxsize= 9 cm 
\leavevmode 
\epsfbox{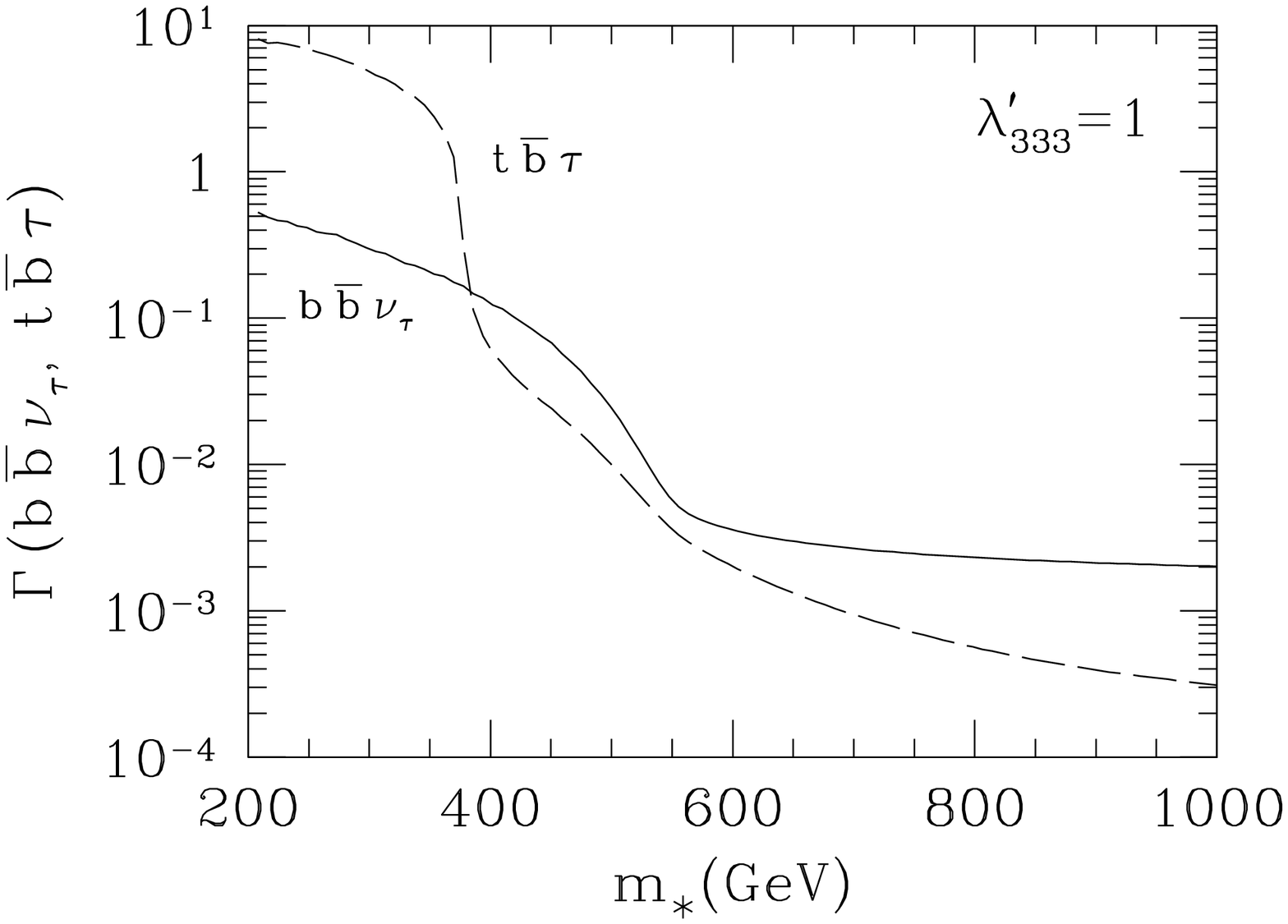} 
\vspace*{0.01truecm}
\epsfxsize= 9 cm 
\epsfbox{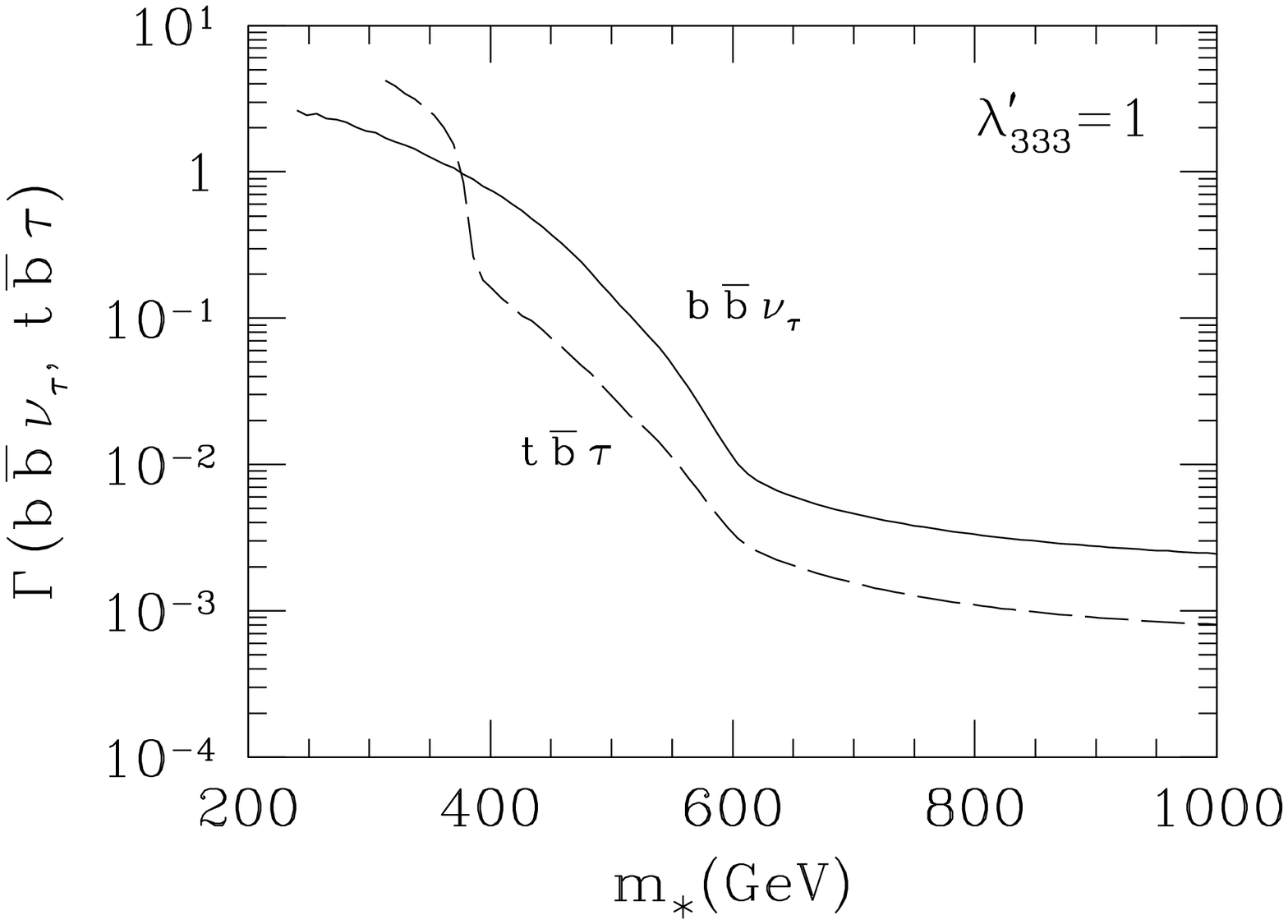} 
\end{center} 
\caption[f1]{$\Gamma(\tilde{\chi}^0_1 \to  b \bar{b}
\nu_{\tau})$ and $\Gamma(\tilde{\chi}^0_1 \to  t \bar{b} {\tau})$
(in\,GeV), respectively solid and long-dashed lines, versus
$m_\ast = m_{\tilde{Q}} = m_{\tilde{U^c}} = m_{\tilde{D^c}}$,
for $\mu =  600\,$GeV, $m_{\tilde{L}} = m_{\tilde{E^c}} 
= 600\,$GeV, $m_{\tilde{B}}=600\,$GeV, and $m_{\tilde{W}}=2
m_{\tilde{B}}$. The value of $\tan \beta$ is $3$ in the upper frame,
$30$ in the lower one.  All trilinear soft terms are fixed at
$350\,$GeV in the upper frame, and at $150\,$GeV in the lower
frame.}
\label{binohino-mix}
\end{figure}

If a large mixing Bino-Higgsino is allowed, then, a substantial
increase in the width of $\Gamma(\tilde{\chi}^0_1\to t \bar{b}
{\tau})$ is expected for low $\tan \beta$ and an increase of
$\Gamma(\tilde{\chi}^0_1\to b\bar{b}\nu_{\tau})$ for large
$\tan\beta$. These features are shown explicitly in
figure~\ref{binohino-mix}, where $\mu$, $m_{\tilde{B}}$ as well as the
doublet and singlet slepton masses are all fixed at $600\,$GeV.  The
value of $\tan \beta =3$ and $A_t=A_b=A_\tau = 350\,$GeV are used in
the upper frame, $\tan \beta =30$ and $A_t=A_b=A_\tau = 150\,$GeV in
the lower frame.  The curves stop when $m_{\tilde{b}} < m_{t}$ in the
case of the width $\Gamma(\tilde{\chi}^0_1\to t \bar{b} {\tau})$, and
$m_{\tilde{b}}< 70\,$GeV (with $70\,$GeV an average cut that should
mimic more complicated and model-dependent experimental lower bounds),
in the case of the width $\Gamma(\tilde{\chi}^0_1\to b
\bar{b}\nu_{\tau})$.

In conclusion, when only the coupling $\lambda^\prime_{333}$ is
nonnegligible, the lightest neutralino decays, in general, as
$\tilde{\chi}^0_1\to b\bar{b}\nu_{\tau}$. A final state with $4b$'s
and missing energy is therefore the distinctive signature of two
pair-produced neutralinos. Excesses of $b$-quarks at hadron collider
may be obtained from sfermions decaying through gauge interactions
into neutralinos and the corresponding fermions.  The two massive
decay modes $\tilde{\chi}^0_1\to t\bar{b}{\tau}$ and
$\tilde{\chi}^0_1\to \bar{t} b \bar{\tau}$ are, in general,
subdominant, although there are wide regions of parameter space in
which they have rates of the same order of magnitude as that of the
decay $\tilde{\chi}^0_1\to b\bar{b}\nu_{\tau}$. The rates for the two
decays into $t$-quarks are, however, the largest when the lightest
neutralino has a large Higgsino component and $\tan \beta$ is not too
large. This dominance is observed in the region of resonant neutralino
decay into an intermediate $\tilde{t}$-squark.

A few more comments are in order here regarding the choice of slepton
masses made for the different phenomenological situations illustrated
in this section.  Fixed values of slepton masses were chosen, smaller
than the largest values of squark masses and, in general, bigger than
the smallest values. In most of the known supersymmetric models,
however, slepton masses are smaller than squark masses and, being
linked to the gravitino mass, as the squark masses, they also increase
when the squark masses increase. If, for example, the choice
$m_{\tilde{L}}=m_{\tilde{E^c}}= (1/2) m_{\tilde{Q}}$
($m_{\tilde{Q}}=m_{\tilde{U^c}}=m_{\tilde{D^c}}$) is made, then the
slepton-exchange diagram becomes more important than it is in the
different plots shown in this section. In general the widths are
enhanced both in the resonant and nonresonant region. In particular,
in the nonresonant region, widths are always decreasing and do not
exhibit the typical plateau that indicates now the slepton-exchange
contribution to be larger than those with exchange of the lightest
squarks.  An enhancement is also expected when the slepton-exchange
contribution is resonant: i.e.\ a similar type of enhancement for both
$b \bar{b}\nu_\tau$ and $t \bar{b}\tau$ modes at small $\tan \beta$,
and possibly a larger enhancement for the $t \bar{b}\tau$ mode at
large $\tan \beta$, or more generically, when large left-right mixing
terms are present in the charged-slepton mass matrix.

Finally, we show how the widths presented so far change when lower
values of the coupling $\lambda^\prime_{333}$ are considered. We
illustrate the dependence of the two widths
$\Gamma(\tilde{\chi}^0_1\to b \bar{b}\nu_{\tau})$ and
$\Gamma(\tilde{\chi}^0_1\to t \bar{b}\tau)$ on this coupling, for the
particular direction of parameter space considered in the upper frame
of figure~\ref{binohino-mix}.

\begin{figure}[ht] 
\begin{center}
\epsfxsize= 9 cm 
\leavevmode 
\epsfbox{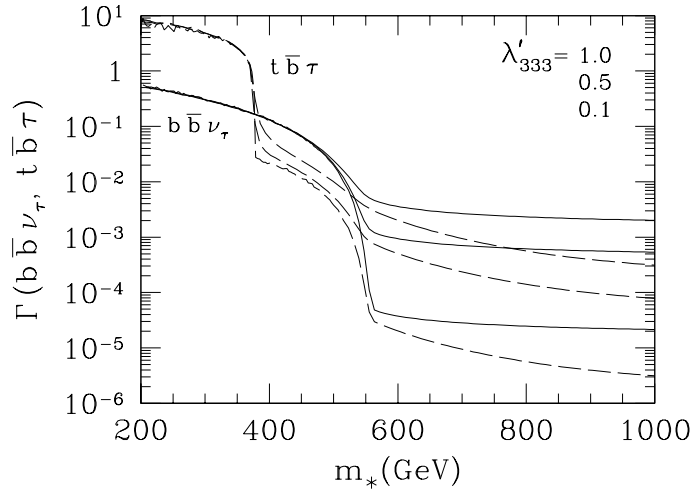}
\end{center} 
\caption{Dependence of
$\Gamma(\tilde{\chi}^0_1 \to  b \bar{b} \nu_{\tau})$ (solid lines)
and $\Gamma(\tilde{\chi}^0_1\to t \bar{b} {\tau})$ (long-dashed lines)
on the coupling $\lambda_{333}^\prime$. All other parameters are
chosen as in the upper frame of figure~\ref{binohino-mix}. The values
of $\lambda_{333}^\prime$ are 1, 0.5, 0.1 from top to
bottom.}
\label{lambdasmall}
\end{figure}

In figure~\ref{lambdasmall}, the two widths are shown in solid
($\Gamma(\tilde{\chi}^0_1\to b \bar{b}\nu_{\tau})$) and dashed
($\Gamma(\tilde{\chi}^0_1\to t \bar{b} \tau)$) lines, for different
values of $\lambda_{333}^\prime$, i.e.\ $\lambda_{333}^\prime = 1$,
0.5 and 0.1.  In the resonant sfermion region, there is very little
dependence on the value of $\lambda_{333}^\prime$ chosen.  Off
resonance, the widths are scaled down by the factor $\vert
\lambda_{333}^\prime \vert^2$.  This feature is rather general and
holds also in the case of other directions of parameter space.

\subsection{$\lambda_{333}^\prime$ and $\lambda_{323}^\prime\ne 0$}
\label{resotherlprime}

It is possible that apart from $\lambda_{333}^\prime$, some other
couplings of type $\lambda^\prime$ are nonnegligible. As discussed in
section~\ref{othercouplings}, couplings such as $\lambda_{233}^\prime$
and $\lambda_{332}^\prime$ give rise, as $\lambda_{333}^\prime$, to
decays of the lightest neutralino into massive and practically
massless final states. Qualitatively, the results obtained for
$\lambda_{333}^\prime$ in the previous subsection, therefore, also
apply to these couplings.

\begin{figure}[ht] 
\begin{center}
\epsfxsize= 9 cm 
\leavevmode 
\epsfbox{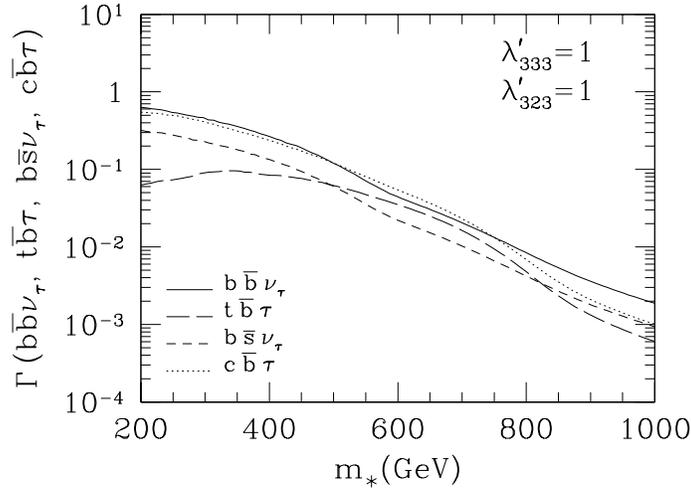}
\end{center} 
\caption{Widths (in\,GeV)
for the four decays $\tilde{\chi}^0_1\to b \bar{b} \nu_{\tau}$, $t
\bar{b} {\tau}$, $b \bar{s} \nu_\tau$, and $c \bar{b}\nu_\tau$, when
only the two couplings $\lambda^\prime_{333}$ and
$\lambda^\prime_{323}$ are nonvanishing, as function of the sfermion
mass $m_\ast$, equal to
$m_{\tilde{Q}}=m_{\tilde{U^c}}=m_{\tilde{D^c}}$, and to $(1.4)\times
m_{\tilde{L}}$ (with $m_{\tilde{L}}=m_{\tilde{E^c}}$), for $\mu = 
1500\,$GeV, $m_{\tilde{B}}=600\,$GeV, and $m_{\tilde{W}}=2
m_{\tilde{B}}$. The value of $\tan \beta$ is $3$ and all trilinear
soft terms are chosen in such a way to have vanishing left-right
mixing terms in the sfermion mass matrices squared.}
\label{lplp}
\end{figure}

Different is the situation obtained for the coupling
$\lambda_{323}^\prime$. As mentioned already, only final states with
light particles can be obtained in this case: $\nu_\tau s \bar{b}$,
$\nu_\tau \bar{s} b$ and $c \bar{b}\tau$, $\bar{c} b
\bar{\tau}$. Those with charged leptons are not phase-space suppressed
as the decay modes $t \bar{b}\tau$ and $\bar{t} b \bar{\tau}$
discussed in the previous subsection. The widths of the four decay
modes $b \bar{b} \nu_\tau$, $t
\bar{b}\tau$, $b \bar{s} \nu_\tau$, and $c \bar{b}\tau$ are shown
explicitly in figure~\ref{lplp}, when both couplings
$\lambda_{333}^\prime$ and $\lambda_{323}^\prime$ are equal to $1$.
They are plotted as functions of the mass of the sfermions virtually
exchanged in the decays. Differently than in the previous figures, the
common slepton mass ($m_{\tilde{L}}=m_{\tilde{E^c}}$), is now varying
together with the common squark mass
($m_{\tilde{Q}}=m_{\tilde{U^c}}=m_{\tilde{D^c}}$) and the ratio
$m_{\tilde{L}}/m_{\tilde{Q}}$ is, for simplicity, fixed to the value
$\simeq 0.7$.  The aim is to visualize, if possible, the dependence on
the $\tilde{\tau}$-slepton mass in the resonant region 
$600 \ltap m_\ast \ltap 860\,$GeV. In this region, two decay modes
are present $\tilde{\tau} \to b \bar{t}$ and $b \bar{c}$, with a
nontrivial dependence on the $\tilde{\tau}$-slepton mass. A mainly Bino
lightest neutralino is considered (the values $m_{\tilde{B}}=600\,$GeV,
$m_{\tilde{W}}=2 m_{\tilde{B}}$, and $\mu =  1500\,$GeV are used) for
this figure. Finally, the choice $\tan \beta=3$ is made, and all
trilinear soft terms are chosen in such a way to have vanishing
left-right terms in all sfermion mass matrices.

\begin{figure}[ht] 
\begin{center}
\epsfxsize= 9 cm 
\leavevmode 
\epsfbox{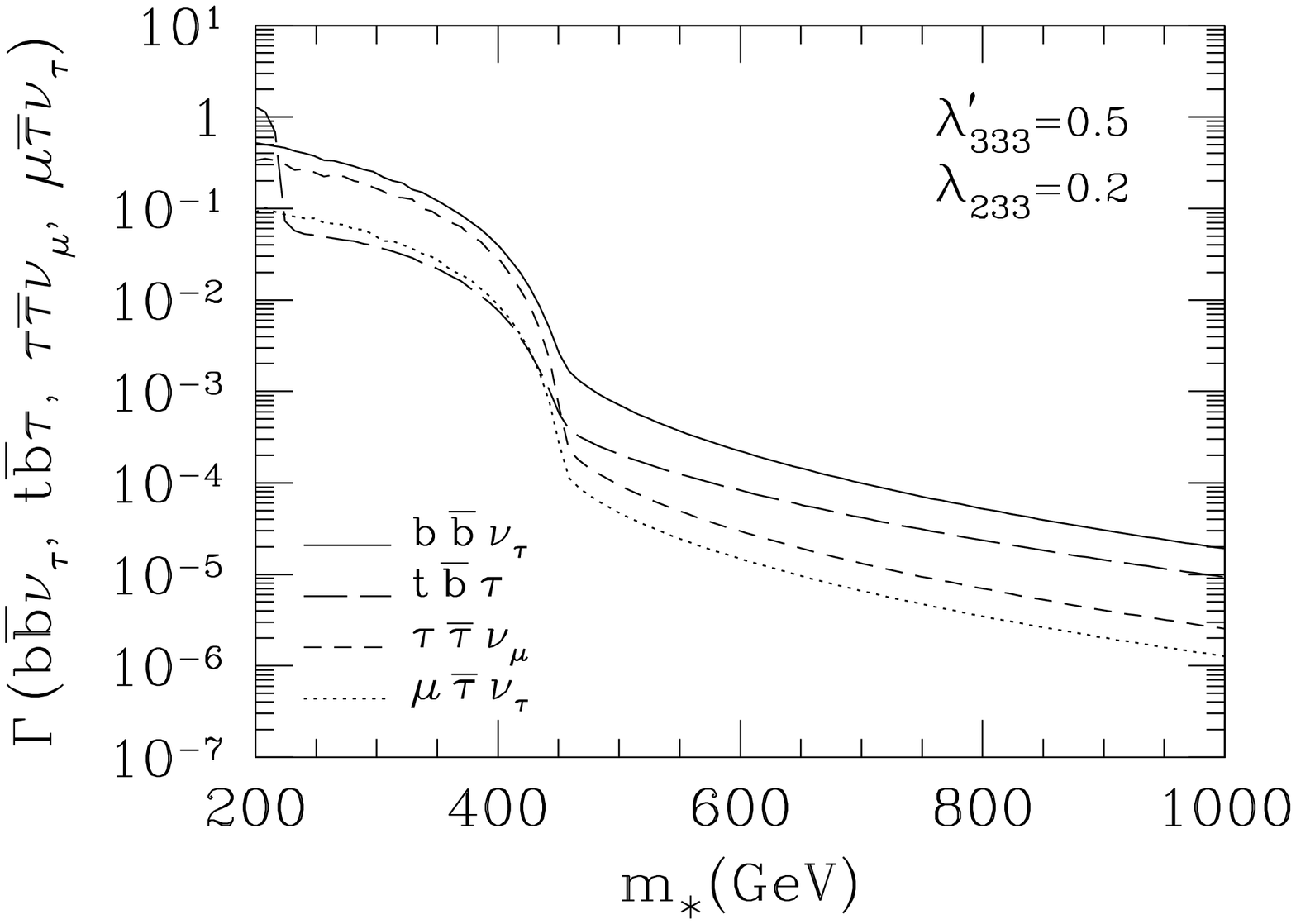} 
\vspace*{0.01truecm}
\epsfxsize= 9 cm 
\epsfbox{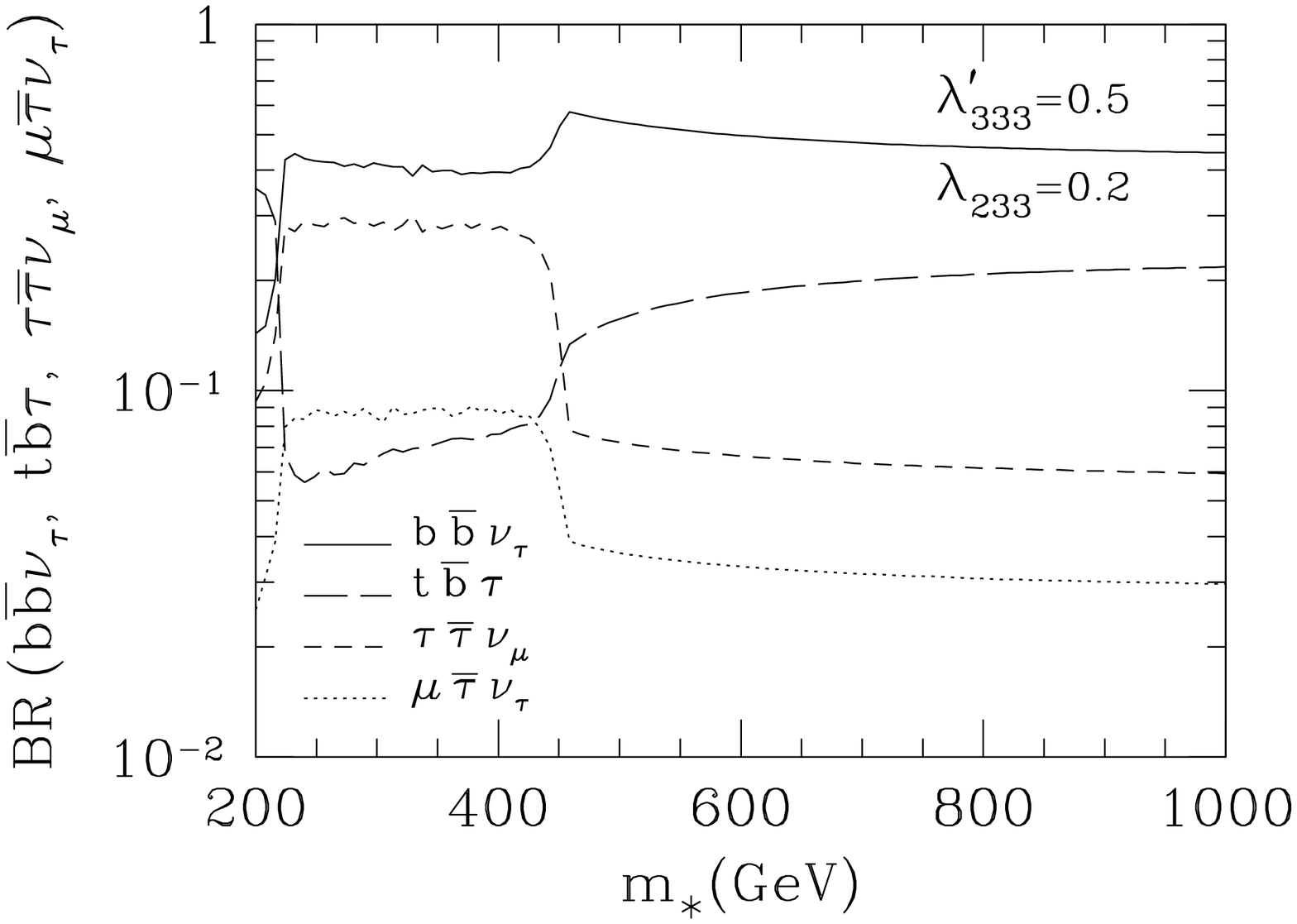} 
\end{center} 
\caption[f1]{Widths (in\,GeV) and
branching ratios for the four decays $\tilde{\chi}^0_1\to b \bar{b}
\nu_{\tau}$, $t \bar{b} {\tau}$, $\tau
\bar{\tau} \nu_\mu$, and $\mu \bar{\tau}\nu_\tau$, when only the two
couplings $\lambda^\prime_{333}$ and $\lambda_{233}$ are nonvanishing,
as function of the sfermion mass $m_\ast$, equal to
$m_{\tilde{Q}}=m_{\tilde{U^c}}=m_{\tilde{D^c}}$, and
$m_{\tilde{L}}=m_{\tilde{E^c}}$, for $\mu =  500\,$GeV,
$m_{\tilde{B}}=500\,$GeV, and $m_{\tilde{W}}=2 m_{\tilde{B}}$. The
value of $\tan \beta$ is $3$ and all trilinear soft terms are chosen
in such a way to have vanishing left-right mixing terms in the
sfermion mass matrices squared.}
\label{landlp}
\end{figure}
 
The distance between the curve relative to $\tilde{\chi}^0_1\to c
\bar{b}\tau$ and $\tilde{\chi}^0_1\to t \bar{b}\tau$ in
figure~\ref{lplp}, is essentially explained by the phase-space
suppression suffered by the decay mode $\tilde{\chi}^0_1\to t
\bar{b}\tau$.  In spite of the more complicated decay possibilities
that the $\tilde{b}$-squark has in the resonant region $m_{\tilde{b}}
< m_{\tilde{\chi}^0_1}$, the decay $\tilde{\chi}^0_1\to b \bar{s}
\nu_\tau$ has a width which is about 1/2 of that for the decay
$\tilde{\chi}^0_1\to b \bar{b} \nu_\tau$ throughout the whole range of
sfermion masses. Notice that the decay mode $\tilde{\chi}^0_1\to b
\bar{b}\nu_\tau$ actually collects the final states $b \bar{b}
\nu_\tau$ and its conjugated one $\bar{b} b \bar{\nu}_\tau$,
which are identical in the approximation of massless neutrinos.
Notice further that the width for the channel $\tilde{\chi}^0_1\to c
\bar{b}\tau$ in the resonant slepton region, is about as large as the
width for the channel $\tilde{\chi}^0_1 \to b \bar{b} \nu_\tau$, in
spite of the fact that the final state $b \bar{b} \nu_\tau$ includes
the state $b \bar{b} \nu_\tau$ and its conjugated one.  This can be
understood as follows. In this region, the slepton-mediated diagrams
are obviously the dominant ones. The lightest neutralino width is then
well approximated, say in the case of the decay into $ c \bar{b}\tau$,
as follows: $\Gamma(\tilde{\chi}^0_1\to c \bar{b}\tau) \simeq
\Gamma(\tilde{\chi}^0_1 \to \tau \tilde{\tau}_1^\ast) \times
BR(\tilde{\tau}_1^\ast \to c \bar{b}) $. In the limit of almost pure
Bino for the lightest neutralino, and in the absence of left-right
mixing terms in the slepton mass matrices, we have
$\Gamma(\tilde{\chi}^0_1 \to \tau \tilde{\tau}_1^\ast) =
\Gamma(\tilde{\chi}^0_1 \to \nu_\tau \tilde{\nu_\tau}) $.  In this
same approximation, and for $\lambda_{333}^\prime
=\lambda_{323}^\prime$, we also have $BR(\tilde{\tau}_1^\ast \to c
\bar{b}) \simeq BR(\tilde{\nu}_\tau \to b \bar{b})$.  Remember that:
$BR(\tilde{\tau}_1^\ast \to c \bar{b}) =
\Gamma_{\tilde{\tau}^\ast}(\lambda_{323}^\prime)/
(\Gamma_{\tilde{\tau}^\ast}(\lambda_{323}^\prime)+
\Gamma_{\tilde{\tau}^\ast}(\lambda_{333}^\prime))\simeq 1/2$ and
$BR(\tilde{\nu}_\tau \to b \bar{b}) =
\Gamma_{\tilde{\nu}_\tau}(\lambda_{323}^\prime)/
(\Gamma_{\tilde{\nu}_\tau}(\lambda_{323}^\prime)+
\Gamma_{\tilde{\nu}_\tau}(\lambda_{333}^\prime))\simeq 1/2$, see
appendix~\ref{sf_widths}. This explain the near equality of
$\Gamma(\tilde{\chi}^0_1 \to c \bar{b}\tau)$ and
$\Gamma(\tilde{\chi}^0_1 \to b \bar{b}\nu_\tau)$, in spite of the fact
that the state $b \bar{b} \nu_\tau$ is self-conjugated.

\subsection{$\lambda_{333}^\prime$ and $\lambda_{233}\ne 0$}
\label{resolprimeandl}

\looseness=1 Let us now assume the somewhat simplified situation in
which only one coupling of type $\lambda^\prime$,
$\lambda_{333}^\prime$, is nonnegligible together with a coupling of
type $\lambda$, i.e.\ $\lambda_{233}$. As anticipated in
section~\ref{othercouplings}, the lightest neutralino decays induced
by $\lambda_{233}$ are $\tilde{\chi}^0_1 \to \tau \bar{\tau} \nu_\mu$,
$\mu
\bar{\tau}\nu_\tau$, and $\bar{\mu} {\tau} \nu_\tau$.

If $\lambda_{333}^\prime$ and $\lambda_{233}$ have the same size and
$\tilde{\chi}^0_1$ is mainly a Bino, the leptonic decay modes, induced
by $\lambda_{233}$ always dominate over those induced by
$\lambda_{333}^\prime$. However, if $\lambda_{233}$ is smaller than
$\lambda_{333}^\prime$, even only by a factor, say, 2 or 3, then the
decay into $b \bar{b} \nu_\tau$, induced by $\lambda_{333}^\prime$
dominates over the leptonic ones, which, in turn, may have larger
widths than the massive decay modes $t \bar{b}\tau$ and $\bar{t} b
\bar{\tau}$.  This situation is, in particular, realized when the
neutralino is a very mixed state with nonnegligible or large Wino and
Higgsino components, as shown in figure~\ref{landlp}.


In this figure, widths and branching ratios obtained for the four
decays, $\tilde{\chi}^0_1\to b \bar{b} \nu_{\tau}$, $t \bar{b}
{\tau}$, $\tau \bar{\tau} \nu_\mu$, and $\mu \bar{\tau}\nu_\tau$, are
shown as function of $m_{\tilde{Q}}=m_{\tilde{U^c}}=m_{\tilde{D^c}}$
when $\lambda_{333}^\prime =0.5$ and $\lambda_{233}=0.2$.  The slepton
masses are assumed to be equal to the squark masses:
$m_{\tilde{Q}}=m_{\tilde{L}}=m_{\tilde{E^c}}$. The gaugino masses are
$m_{\tilde{B}}= (1/2) m_{\tilde{W}} =500\,$GeV, and $500\,$GeV is also
the value assigned to $\mu$.  The value of $\tan \beta$ is fixed at 3,
and the trilinear $A$ terms are chosen in such a way to have vanishing
left-right mixing terms in all sfermion mass matrices.  Branching
ratios are defined as ratios of the width for the different decay
modes over the total width for the lightest neutralino. This, in turn,
is given by:
\begin{equation} 
 \Gamma_{tot}(\tilde{\chi}^0_1)  =
 \Gamma(b \bar{b} \nu_{\tau})    +  2\Gamma(t \bar{b} {\tau}) + 
 \Gamma(\tau \bar{\tau} \nu_\mu) + 2\Gamma(\mu \bar{\tau}\nu_\tau)\,.
\label{totalwidth}
\end{equation}
As the figure clearly shows, all widths and branching ratios are not
too dissimilar, for the choice of $\lambda_{333}^\prime$ and
$\lambda_{233}$ made here.  Notice how the scaling factor of 2 between
the curves for $\Gamma(\tilde{\chi}^0_1\to \tau \bar{\tau}\nu_{\mu})$
and $\Gamma(\tilde{\chi}^0_1\to \mu \bar{\tau}\nu_{\tau})$, which
would be expected for an almost pure Bino lightest neutralino, is here
reduced by the substantial Higgsino and smaller Wino component.

\section{Summary and conclusions}
\label{concl}

The instability of the LSP is, perhaps, the most dramatic feature of
$R_p$-violating models. Whether it is the LSP or not, the lightest
neutralino decays through $R_p$-violating interactions. Thanks to
$R_p$-violating bilinear terms in the superpotential, the lightest
neutralino has leptonic components. These leptonic admixtures are, in
general, very small, since directly constrained by the smallness of
neutrino masses. Among the other three $R_p$-violating terms in the
superpotential, two violate explicitly the lepton number, $L$,
$\lambda^\prime L Q D^c$ and $\lambda L L E^c$, and the third,
$\lambda'' U^c D^c D^c$, violates the baryon number, $B$. Their
simultaneous presence induces a too rapid proton decay, which is
experimentally excluded. It is, then, in general assumed that one of
these two quantum numbers is still a conserved one, whereas
$R_p$-violation is induced by the nonconservation of the second one.

\looseness=1 The working assumption in this paper is that $B$ remains
the conserved quantum number. It is the violation of $L$, then, that
is responsible for the striking phenomenology of $R_p$-violating
models, and that induces, together with gauge interactions, the decay
of the lightest neutralino. $L$ violation may also give rise to
unusual signatures for slepton decays directly, through $\tilde{l} \to
q^\prime \bar{q}$, or indirectly, through $L$-violating decays of the
lightest neutralino, and/or chargino, and to unusual production
mechanisms for sleptons at colliders. Indeed, it was recently pointed
out that the coupling $\lambda^\prime_{333}$ may be responsible for
the single production of $\tilde{\tau}$-sleptons at the LHC, through
the processes $pp\to t
\bar{b}\tilde{\tau}X$, $pp\to t \tilde{\tau}X$, with cross sections
comparable to those for the corresponding single-production processes
of charged Higgs bosons, $pp\to t \bar{b}H^-X$, $pp\to t
H^-X$~\cite{Borzumati:1999th}. (Single production of charged sleptons
and charged Higgs bosons at the Tevatron, although possible, is
certainly more difficult to detect than at the LHC.) That the coupling
$\lambda^\prime_{333}$ may be large, or simply the dominant among
other $R_p$-violating couplings, is a possibility that has not yet
been challenged by experiment. Perhaps, negative searches for the
abovementioned production processes and the combined
decays for the $\tilde{\tau}$-sleptons may provide in the future
severe restrictions in a region of the supersymmetric parameter space
of which $\lambda^\prime_{333}$ is one coordinate. For now, however,
if such a coupling is the dominant one, it may be the main source of
neutralino decays.  The distinguishing property of this coupling with
respect to other $L$-violating couplings is that it involves the
$t$-quark in association with a $\tau$-lepton. If heavy enough, the
lightest neutralino can therefore give rise to a very massive final
state. Such a possibility was never considered before, assuming
perhaps that decays of $\tilde{\chi}^0_1$ into massive particles would
be handicapped by a severe phase space suppression.

In this paper, neutralino decays are studied without any restrictive
assumptions as to whether the final state is massive or massless. The
coupling $\lambda^\prime_{333}$ offers a good ground to study both
possibilities and to understand under which conditions the massive
decay mode may be sizable with respect to the massless one. It is also
representative of other couplings, such as $\lambda^\prime_{233}$ or
$\lambda^\prime_{332}$, that may also give rise to final states
containing the $t$-quark. Given the complexity of phenomenological
analyses in which many $L$-violating couplings are simultaneously
present, some particular choices are made. First, it is considered the
case in which one coupling only, i.e.\ $\lambda^\prime_{333}$, is
largely dominating over the others, which can be neglected.  Later, it
is assumed that two couplings at a time are dominant over the other
ones, as for example $\lambda^\prime_{333}$ and
$\lambda^\prime_{323}$, or $\lambda^\prime_{333}$ and $\lambda_{233}$,
with $\lambda^\prime_{333}$ the only one of the two capable of
inducing a massive final state.

It should be remarked that the importance of the coupling
$\lambda^\prime_{333}$, and of other couplings of the same type,
$\lambda^\prime_{323}$, $\lambda^\prime_{332}$, etc is also due to
fact that, being potentially large, they may pollute the signals at
incoming colliders for production and decays of charged Higgs
bosons. Through these couplings, a $\tilde{\tau}$-slepton singly
produced in association with a $t$- and a $b$-quark can easily mimic a
charged Higgs boson giving rise to the same final states.  It is
therefore important to discern all possible consequences that these
couplings may have. A detailed knowledge of the decays of
$\tilde{\tau}$'s induced directly by these couplings or indirectly,
through decays of charginos and neutralinos may be crucial to reach an
unambiguous identification of scalar particles at future colliders.

\looseness=1 The present analysis has to be considered as a first step
of a more complete program of sparticle-decay studies in
$R_p$-violating models~\cite{PREP}. Rather simple and general
analytical expressions for neutralino decays into massless or massive
final fermions are given. They are easily adaptable to any
$R_p$-violating couplings, and to heavier neutralinos. They include
the possibility of both resonant and nonresonant decays, the former
one describing the situation in which the lightest neutralino is not
the LSP.  Some of these $3$-body decays were previously discussed in
the literature and some were also incorporated in event generators
used in previous (LEP) and forthcoming (Tevatron) experimental
analyses. For these cases, the present study should provide useful
cross checks and guidelines to identify the directions of parameter
space in which the $R_p$-violating signals are more prominent. (We
emphasize here again that experimental analyses based on a scan of the
parameter space obtained through Renormalization Group Equations from
only 5 high-scale parameters, $m$, $\mu$, $M$, $A_0$, $B$, as in the
mSUGRA model, are not valid when trying to obtain limit on not too
small $R_p$-violating couplings.)  Analyses of decays into
purely-third-generations fermions, including the $t$-quark are
genuinely new.  Analytical expressions for these decays are given in a
form that can be easily implemented in event generators.\footnote{The
FORTRAN coded expressions for the amplitudes squared are available
upon request.}

\looseness=1 The results may be summarized as follows.  Quite
generically, if the lightest neutralino is lighter (or only slightly
heavier) than the $t$-quark, it decays predominantly into massless
fermions. Depending on the relative size of the $R_p$-violating
couplings, the dominant decays can be, e.g., $\tilde{\chi}^0_1 \to b
\bar{b}\nu_\tau$, due to a dominant $\lambda_{333}^{\prime}$ coupling,
$\tilde{\chi}^0_1 \to c
\bar{b} \tau$, $\tilde{\chi}^0_1 \to s \bar{b} \nu_i$ and their
CP-conjugate states, induced by $\lambda_{323}^{\prime}$, or
$\tilde{\chi}^0_1 \to \mu \bar{\tau}\nu_\tau$, and $\tilde{\chi}^0_1
\to \bar{\mu} {\tau} \nu_\tau$, due to a sizable or moderate
$\lambda_{233}$ coupling.  When two such couplings are simultaneously
present and of the same size, the rates for various massless states
are of the same order of magnitude for a wide range of parameter
values, see for example figure~\ref{lplp}. The same holds when
$\lambda^\prime_{333}$ and $\lambda_{233}$ are dominating over the
other couplings, see figure~\ref{landlp}.

On the other hand, one can envisage a situation in which the $\tilde
\chi_1^0$ is substantially heavy, so that its decays into final states
containing the $t$-quark are possible, as for example $t \bar{b}
\tau$, $t \bar{s} \tau$, $t \bar{b} \mu$ and their CP conjugated
states.  Massive decay modes imply the virtual exchange of sfermions
of third generations ($\tilde{t}$'s, $\tilde{b}$'s, and
$\tilde{\tau}$'s) that can be relatively light if the left-right
sfermion mixing is large. In this case, decays with the $t$-quark in
the final state can become competitive. In general, for moderate/large
left-right mixing and/or substantial Bino-Higgsino mixing, these decay
modes can be large at not too large values of $\tan \beta$.  Although
in general subdominant, their rate can be comparable to those of
massless modes, see figures~\ref{binohino-mix},~\ref{lplp},
and~\ref{landlp}.  When the $t$-quark does not decay leptonically,
these decay modes of a pair of lightest neutralinos, give rise to the
like-sign dilepton signal typical of $R_p$-violating models.

Further, an overall increase in the total decay width of
$\tilde\chi_1^0$ is observed, if $\tilde\chi_1^0$ is of Wino type, as
it may happen if gaugino mass unification is not imposed, or in
anomaly-mediated supersymmetry-breaking
scenarios~\cite{Gherghetta:1999sw}, or in grand-unification models
with an additional strong hypercolor
group~\cite{NEWUNIF,NEWGAUGINOREL}.

Given the size of the widths obtained in this analysis, in general,
fast decays of the lightest neutralino are expected. If large, the
couplings considered here will undoubtedly play an important role in
future collider searches. Their impact will be significant also for
production mechanisms and decays of squarks and sleptons, as well as
chargino decays. Such studies will be the subject of future
work~\cite{PREP}.

\vspace*{0.5truecm}
\noindent 
{\bf Acknowledgements} \\ 

We thank K.~Hamaguchi, G.~Moultaka, T.~Nagano, and M.~Yamaguchi for
discussions and A.~Arhrib for spotting a typo in the appendix. This
work was supported in part by the Japanese Ministry of Education and
by CNRS.  F.~B. and F.~T. thank the KEK theory group, in particular
Y.~Okada, for hospitality.  F.~B. and R.~G. also acknowledge the
hospitality of the theory division at CERN.

\appendix 
 
\section{Neutralino mass matrix} 
\label{neutralino} 

In the limit of small $k_i$ in eq.~(\ref{superpot}) and small
\emph{vev}'s for the neutral component of $\tilde{L}_i$, the
neutralino mass matrix reduces to the ordinary $4\times 4$ matrix of
$R_p$-conserving models. Thus, on the basis
($\tilde{B},\tilde{W}_3,\tilde{H}_d,\tilde{H}_u,$) the neutralino mass
matrix has the form:
\begin{equation}
 M_{neutr}  =  \pmatrix{
   M_{\tilde{B}} & 0 & -m_Z s_W c_\beta &  m_Z s_W s_\beta   \cr
   0 & M_{\tilde{W}} & m_Z c_W c_\beta  &  -m_Z c_W s_\beta  \cr
   -m_Z s_W c_\beta  & m_Z c_W c_\beta  &   0  & -\mu        \cr
    m_Z s_W s_\beta  & -m_Z c_W s_\beta & -\mu &  0  }.
\end{equation}
Notice that his matrix is symmetric, but in general not hermitian, 
if $M_{\tilde{B}}$, $M_{\tilde{W}}$ and $\mu$ have phases. Therefore, 
the mass eigenvalues are, in general, complex. This inconvenience
can be avoided by a rotation of the corresponding eigenstates. 
Thus, mass eigenstates with real and positive 
eigenvalues are defined as: 
\begin{eqnarray}
  {\tilde{\chi}^0}_i = \eta_i \sum_{\alpha} O_{i\alpha}\psi_\alpha\,,
\end{eqnarray}
where $\psi_a$ are the states $\psi_\alpha = \tilde{B},\tilde{W}_3,
\tilde{H}_d, \tilde{H}_u $ and $\eta_i$ are phase factors.  If
$M_{\tilde{B}}$, $M_{\tilde{W}}$ and $\mu$ are real parameters, the
matrix $ M_{neutr}$ is orthogonal and the mass eigenvalues are real,
with positive and/or negative sign. If needed, positive mass
eigenvalues are obtained still through a rotation, as before, and the
factors $\eta_i$ are $\pm i$ or $\pm 1$.

\section{Scalar superpartner mass and mixing} 
\label{sfermion} 

The $2 \times 2$ squark or slepton mass squared matrix, 
\begin{equation} 
{\cal M}_f^2 = 
\pmatrix{
     m^2_{f\,RR}         & (m^2_{f\,LR})    \cr
     (m^2_{f\,LR})^\ast     &  m^2_{f\,LL} },
\label{massmatr} 
\end{equation} 
written here in the basis $\{ \widetilde{f}_L,\widetilde{f}_R\}$, is
hermitean ${\cal M}_f^2 = ({\cal M}_f^2)^{\dagger}$, with real
eigenvalues
\begin{equation} 
 m^{2}_{\widetilde{f}_{1,2}} = 
 \frac{1}{2} \left\{ \left(m^2_{f\,LL} +m^2_{f\,RR}\right) \mp 
               \sqrt{\left(m^2_{f\,LL} -m^2_{f\,RR}\right)^2 
   +4|m^2_{f\,LR}|^2} 
             \right\}. 
\label{eigenvalues} 
\end{equation}

The eigenvectors $\widetilde{f}_1$ and $\widetilde{f}_2$ 
corresponding to the 
eigenvalues in~(\ref{eigenvalues}) are obtained from $\widetilde{f}_L$, 
$\widetilde{f}_R$ through a unitary transformation : 
\begin{equation} 
\pmatrix{ \widetilde{f}_1 \cr \widetilde{f}_2 }
 =  U_f 
\pmatrix{\widetilde{f}_R \cr \widetilde{f}_L }
\equiv 
\pmatrix{
 \cos \theta_f                & \sin \theta_f e^{i\phi_f}   \cr
 - \sin \theta_f e^{-i \phi_f}  & \cos \theta_f  }
\pmatrix{  \widetilde{f}_R \cr \widetilde{f}_L }, 
\label{urotation} 
\end{equation} 
where $\phi_f = {\rm Arg}(m_{LR}^2)$.  
The mixing angle $\theta_f $ is defined, up to a two-fold ambiguity, 
by the relations 
\begin{equation} 
 \sin 2 \theta_f = 
  -\frac{2 m^2_{f\,LR}}
        {m^2_{\widetilde{f}_2} -m^{2}_{\widetilde{f}_1}} \,; 
\qquad\qquad
 \cos 2 \theta_f = 
  \frac{m^2_{f\,LL} - m^2_{f\,RR}}
        {m^2_{\widetilde{f}_2} -m^2_{\widetilde{f}_1}} \,. 
\end{equation} 

Notice that the entries in the sfermion mass matrix in
eq.~(\ref{massmatr}) contain in general terms proportional to
$R_p$-violating couplings. However, in the limit of small $k_i$ in
eq.~(\ref{superpot}) and small \emph{vev}'s for the neutral component
of $\tilde{L}_i$, $m^2_{f\,RR}$, $m^2_{f\,LL}$, and $(m^2_{f\,LR})$
reduce to the usual entries present in $R_p$-conserving models.

\section{Sfermion widths} 
\label{sf_widths} 

If sufficiently light, the sfermions mediating the neutralino decays
discussed in sections~\ref{lambdaprime333} and~\ref{othercouplings}
may be on shell. The form of the propagators to use in this case is
given in eq.~(\ref{widthpropagator}).

\subsection{Only $\lambda_{333}^\prime\ne 0$}

We consider first the case in which only the coupling
$\lambda^\prime_{333}$ is nonvanishing. In this approximations, there
is only one possible decay mode for the lightest $\tilde{\tau}$
eigenstate $\tilde{\tau}_1$, i.e.\ that induced by the coupling
$\lambda^\prime_{333}$: $\tilde{\tau}_1 \to \bar{t} b$.  (Decay modes
mediated by gauge interactions, such as $\tilde{\tau}_1 \to \tau
{\tilde{\chi}^0}_1$ and $\tilde{\tau}_1 \to \nu_\tau
{\tilde{\chi}^-}_1$, with off-shell ${\tilde{\chi}^0}_1$ and
${\tilde{\chi}^-}_1$ would give rise to subdominant $4$-body decays.)
Thus, it is:
\begin{equation}
 \Gamma_{\tilde{\tau}_1}(\lambda^\prime_{333}) =  
 \Gamma(\tilde{\tau}_1 \to \bar{t}b) =  
 \frac{3}{16 \pi}\vert\lambda^\prime_{333} \vert^2 
  m_{\tilde{\tau}_1} \sin^2 \theta_\tau 
 \left( 1 - \frac{m_b^2}{m_{\tilde{\tau}_1}^2}  
           - \frac{m_t^2}{m_{\tilde{\tau}_1}^2} \right) 
 K^{1/2} \left( 1,\frac{m_b}{m_{\tilde{\tau}_1}},
                    \frac{m_t}{m_{\tilde{\tau}_1}} \right),
\label{wstaulambdaprime}
\end{equation}
where $K$ is the K\"allen function $K(x,y,z) = ( ( x^2-y^2-z^2)^2 - 4
y^2 z^2)$.

Similarly, the width for the decay mode of a third generation
sneutrino, $\tilde{\nu}_\tau \to \bar{b} b$, is given by:
\begin{equation}
 \Gamma_{\tilde{\nu}_\tau}(\lambda^\prime_{333}) =   
   \Gamma \left(\tilde{\nu}_\tau \to \overline{b_L} \,b_R \right) + 
   \Gamma \left(\tilde{\nu}_\tau \to \overline{b_R} \,b_L \right) = 
2\,\Gamma \left(\tilde{\nu}_\tau \to \overline{b_L} \,b_R \right) ,  
\end{equation}
with 
\begin{equation}
 \Gamma(\tilde{\nu}_{\tau} \to \overline{b_L}\,b_R) =  
 \frac{3}{16 \pi}\vert\lambda^\prime_{333} \vert^2 \, 
  m_{\tilde{\nu}_\tau} 
 \left(1 -2 \frac{m_b^2}{m_{\tilde{\nu}_\tau}^2} \right) 
 K^{1/2}\left(1,\frac{m_b}{m_{\tilde{\nu}_\tau}},
                    \frac{m_b}{m_{\tilde{\nu}_\tau}} \right).
\end{equation}

The width for the sbottom squark is obtained as sum of the widths for
the two decay modes $\tilde{b}_1 \to b \nu_\tau$ and 
$\tilde{b}_1 \to t \bar{\tau}$, with the second one present only in the
case $m_{\tilde{\chi}^0_1} > m_{\tilde{b}_1} > m_t$:
\begin{equation}
 \Gamma_{\tilde{b}_1}(\lambda^\prime_{333}) =   
 \Gamma \left(\tilde{b}_1 \to b \nu_\tau \right) + 
 \Gamma \left(\tilde{b}_1 \to t  \bar{\tau} \right) \,.
\end{equation}
The two partial widths are respectively: 
\begin{eqnarray}
\Gamma \left(\tilde{b}_1 \to b \nu_\tau \right)
 & = & 
 \frac{1}{16 \pi}\vert\lambda^\prime_{333} \vert^2 \, 
  m_{\tilde{b}_1}  
 \left(\!1 -\!\frac{m_b^2}{m_{\tilde{b}_1}^2} \right)
  K^{1/2}\!\left( 1,\frac{m_b}{m_{\tilde{b}_1}},0 \right)\,,
\nonumber \\
\Gamma \left(\tilde{b}_1 \to t  \bar{\tau} \right)
 & = & 
 \frac{1}{16 \pi}\vert\lambda^\prime_{333} \vert^2 \, 
  m_{\tilde{b}_1}  \cos^2 \theta_b  
 \left(1 -\frac{m_t^2}{m_{\tilde{b}_1}^2} 
           -\frac{m_\tau^2}{m_{\tilde{b}_1}^2} \right)
  K^{1/2} \left(1,\frac{m_t}{m_{\tilde{b}_1}},
                      \frac{m_\tau}{m_{\tilde{b}_1}} \right),\qquad
\end{eqnarray}
where the first is the width for the two decays $\tilde{b}_1 \to b_L
\nu_\tau$ and $\tilde{b}_1 \to b_R \bar{nu}_\tau$.

Finally, the width for the lightest $\tilde{t}$ eigenstate,
$\tilde{t}_1$:
\begin{equation}
 \Gamma_{\tilde{t}_1}(\lambda^\prime_{333}) =  
 \Gamma({\tilde{t}_1 \to \bar{\tau} b}) =  
 \frac{1}{16 \pi}\vert\lambda^\prime_{333} \vert^2 \, 
  m_{\tilde{t}_1}  \sin^2 \theta_t  
 \left(1 -\frac{m_b^2}{m_{\tilde{t}_1}^2} 
           -\frac{m_\tau^2}{m_{\tilde{t}_1}^2} \right)
  K^{1/2}\left(1,\frac{m_b}{m_{\tilde{t}_1}},
                     \frac{m_\tau}{m_{\tilde{t}_1}} \right),
\end{equation}
is the width of the decay $\tilde{t}_1 \to \bar{\tau} b$. 


Except for the sneutrino case, for which there is only one eigenstate
per generation, the widths given above refer to the lightest
$\tilde{\tau}$, $\tilde{b}$, and $\tilde{t}$ eigenstates. In the less
likely case that any of the heaviest of these sfermion eigenstates,
i.e.\ $\tilde{\tau}_2$, $\tilde{b}_2$, and $\tilde{t}_2$ is also
lighter than ${\tilde{\chi}^0_1}$, the relevant widths can be obtained
by interchanging $\cos \theta_f$ with $\sin \theta_f$ in the above
formulas.

\subsection{$\lambda_{333}^\prime$ and $\lambda_{323}^\prime\ne 0$}

Generalization to the case in which other $\lambda^\prime$ couplings
are nonvanishing, such as $\lambda^\prime_{233}$,
$\lambda^\prime_{323}$, $\lambda^\prime_{332}$, etc., are
straightforward.  In particular, when two such couplings are present,
for example $\lambda'_{333}$ and $\lambda'_{323}$, the
$\tilde{\tau}$-slepton may decay also into light particles, as
$\tilde{\tau}\to \bar{c}b$. The corresponding with is:
\begin{equation} 
 \Gamma_{\tilde{\tau}_1}(\lambda^\prime_{323}) =  
 \Gamma(\tilde{\tau}_1\to \bar{c}b) =
 \frac{3}{16\pi}\vert\lambda'_{323}\vert^2\, 
  m_{\tilde{\tau}_1} \sin^2\theta_{\tau}
 \left(1 -\frac{m_b^2}{m^2_{\tilde{\tau}_1}}
           -\frac{m_{c}^2}{m^2_{\tilde{\tau}_1}}\right)
 K^{1/2}\left(1,\frac{m_b}{m_{\tilde{\tau}_1}},
                    \frac{m_{c}}{m_{\tilde{\tau}_1}}\right),
\end{equation} 
and the total with is given by the sum of 
$ \Gamma_{\tilde{\tau}_1}(\lambda^\prime_{323})$ and 
$\Gamma_{\tilde{\tau}_1}(\lambda^\prime_{333})$. 

The sneutrino $\tilde{\nu}_{\tau}$ can now decay as
$\tilde{\nu}_{\tau}\to \overline{s}b$ and $\tilde{\nu}_{\tau}\to
\overline{b}s$, with the $b$-quark always right-handed. These decays
have identical partial widths,
\begin{equation} 
 \Gamma(\tilde{\nu}_{\tau}\to \overline{s}\,b) =
 \Gamma(\tilde{\nu}_{\tau}\to \overline{b}\,s) =
 \frac{3}{16\pi}\vert\lambda'_{323}\vert^2m_{\tilde{\nu}_{\tau}}
 \left(1 -\frac{m_s^2}{m^2_{\tilde{\nu}_{\tau}}}
           -\frac{m_b^2}{m^2_{\tilde{\nu}_{\tau}}}\right)
 K^{1/2}\left(1,\frac{m_s}{m_{\tilde{\nu}_{\tau}}},
                    \frac{m_b}{m_{\tilde{\nu}_{\tau}}}\right).
\end{equation} 
The width due to the coupling $\lambda^\prime_{323}$, is then given by 
\begin{equation}
 \Gamma_{\tilde{\nu}_\tau}(\lambda^\prime_{323}) =   
   \Gamma \left(\tilde{\nu}_\tau \to \overline{s} \,b \right) + 
   \Gamma \left(\tilde{\nu}_\tau \to \overline{b} \,s \right) = 
2\,\Gamma \left(\tilde{\nu}_\tau \to \overline{s} \,b \right).  
\end{equation}
The total width, when both couplings $\lambda^\prime_{333}$ and
$\lambda^\prime_{323}$ are present, is obtained by summing
$\Gamma_{\tilde{\nu}_\tau}(\lambda^\prime_{333})$ and
$\Gamma_{\tilde{\nu}_\tau}(\lambda^\prime_{323})$.

Because of the coupling $\lambda^\prime_{323}$, the $\tilde{b}$-squark
can now decay into light particles, as $\tilde{b}_1\to c\bar{\tau}$,
and $\tilde{b}_1\to s\nu_{\tau}$. The corresponding widths are:
\begin{eqnarray}
 \Gamma(\tilde{b}_1\to c\bar{\tau}) &=& 
 \frac{1}{16\pi}\vert\lambda'_{323}\vert^2\,
  m_{\tilde{b}_1} \cos^2\theta_b
 \left(1 -\frac{m_c^2}{m^2_{\tilde{b}_1}}
           -\frac{m_{\tau}^2}{m^2_{\tilde{b}_1}}\right)
 K^{1/2}\left(1,\frac{m_c}{m_{\tilde{b}_1}},
                    \frac{m_{\tau}}{m_{\tilde{b}_1}}\right)\qquad
\\
 \Gamma(\tilde{b}_1\to s\nu_{\tau}) &=& 
 \frac{1}{16\pi}\vert\lambda'_{323}\vert^2 \,
  m_{\tilde{b}_1} \cos^2\theta_b
 \left(1 -\frac{m_s^2}{m^2_{\tilde{b}_1}}\right)
  K^{1/2}\left(1,\frac{m_s}{m_{\tilde{b}_1}}, 0\right) .
\label{wsblp}
\end{eqnarray}
The total width $\Gamma_{\tilde{b}_1}$ is given by the sum of 
$\Gamma_{\tilde{b}_1}(\lambda^\prime_{333})$ and 
$\Gamma_{\tilde{b}_1}(\lambda^\prime_{323})$, where   
\begin{equation}
 \Gamma_{\tilde{b}_1}(\lambda^\prime_{323}) =   
 \Gamma \left(\tilde{b}_1 \to s \nu_\tau \right) + 
 \Gamma \left(\tilde{b}_1 \to c  \bar{\tau} \right).
\end{equation}

Finally, the $\tilde{c}$- and $\tilde{s}$-squarks decay as
$\tilde{c}_1\to \bar{\tau}b$ and $\tilde{s}_1\to b\nu_{\tau}$,
respectively, with widths:
\begin{eqnarray}
 \Gamma_{\tilde{c}_1}(\lambda^\prime_{323}) &=&  
 \Gamma(\tilde{c}_1\to \bar{\tau}b) = 
 \frac{1}{16\pi} \vert\lambda'_{323}\vert^2 \,
  m_{\tilde{c}_1} \sin^2\theta_c
 \left(1 -\frac{m_b^2}{m^2_{\tilde{c}_1}}
           -\frac{m_{\tau}^2}{m^2_{\tilde{c}_1}} \right)
 K^{1/2}\left(1,\frac{m_b}{m_{\tilde{c}_1}},
                    \frac{m_{\tau}}{m_{\tilde{c}_1}}\right),
\nonumber\\ \\
 \Gamma_{\tilde{s}_1}(\lambda^\prime_{323}) &=&  
 \Gamma(\tilde{s}_1\to b\nu_{\tau}) =
 \frac{1}{16\pi} \vert\lambda'_{323}\vert^2 \,
   m_{\tilde{s}_1} \sin^2\theta_s
 \left(1 -\frac{m_b^2}{m^2_{\tilde{s}_1}}\right)
  K^{1/2}\left(1,\frac{m_b}{m_{\tilde{s}_1}},  0\right)
\end{eqnarray}

\subsection{$\lambda_{333}^\prime$ and $\lambda_{233}\ne 0$}

If besides $\lambda_{333}^\prime$, also $\lambda_{233}$ is
nonvanishing, then, $\tilde{\tau}_1$ can also decay into $\tau
\nu_\mu$ and $\mu \nu_{\tau}$, with corresponding widths:
\begin{eqnarray}
 \Gamma(\tilde{\tau}_1 \to \tau \nu_\mu)
 & = &   
 \frac{1}{16 \pi}\vert\lambda_{233} \vert^2 \, 
  m_{\tilde{\tau}_1}  
 \left(1 -\frac{m_{\tau}^2}{m_{\tilde{\tau}_1}^2} \right) 
  K^{1/2}\left(1,\frac{m_{\tau}}{m_{\tilde{\tau}_1}},0 \right),
\nonumber \\
 \Gamma(\tilde{\tau}_1 \to \bar{\nu}_\tau \mu)
 & = &   
 \frac{1}{16 \pi}\vert\lambda_{233} \vert^2 \, 
  m_{\tilde{\tau}_1} \sin^2 \theta_\tau 
 \left(1 -\frac{m_\mu^2}{m^2_{\tilde{\tau}_1}}\right)
  K^{1/2}\left(1,\frac{m_\mu}{m_{\tilde{\tau}_1}},  0\right),
\label{wstaulambda}
\end{eqnarray}
where the first is the width for $\tilde{\tau}_1 \to \tau_L \nu_\mu$
and $\tilde{\tau}_1 \to \tau_R \bar{nu}_\mu$.  After defining:
\begin{equation}
 \Gamma_{\tilde{\tau}_1}(\lambda_{233}) =   
 \Gamma \left(\tilde{\tau}_1 \to \tau \nu_\mu \right) + 
 \Gamma \left(\tilde{\tau}_1 \to \tau \nu_\tau\right) .
\end{equation}
the total width for $\tau_1$, $\Gamma_{\tilde{\tau}_1} $, is obtained
summing up $\Gamma_{\tilde{\tau}_1}(\lambda_{233})$ to the width
$\Gamma_{\tilde{t}_1}(\lambda^\prime_{333})$ in
eq.~(\ref{wstaulambdaprime}).

Similarly, $\tilde{\nu}_\tau$ can decay into $\bar{\mu}\tau$ and
$\bar{\tau}\mu$:
\begin{equation}
 \Gamma(\tilde{\nu}_{\tau} \to \bar{\mu}  \tau) =  
 \Gamma(\tilde{\nu}_{\tau} \to \bar{\tau} \mu)  =  
 \frac{1}{16 \pi}\vert\lambda^\prime_{333} \vert^2 \, 
  m_{\tilde{\nu}_\tau} 
 \left(1 -\frac{m_\tau^2}{m_{\tilde{\nu}_\tau}^2}
           -\frac{m_\mu^2}{m_{\tilde{\nu}_\tau}^2} \right) 
 K^{1/2}\left(1,\frac{m_\tau}{m_{\tilde{\nu}_\tau}},
                    \frac{m_\mu}{m_{\tilde{\nu}_\tau}}\right),
\end{equation}
and 
\begin{equation}
 \Gamma_{\tilde{\nu}_\tau}(\lambda_{233}) =   
 \Gamma(\tilde{\nu}_{\tau} \to \bar{\mu}  \tau) +  
 \Gamma(\tilde{\nu}_{\tau} \to \bar{\tau} \mu)  \,.  
\end{equation}
Also in this case, the total width $\Gamma_{\tilde{\nu}_\tau}$ is
obtained summing up $\Gamma_{\tilde{\nu}_\tau}(\lambda_{233})$ and
$\Gamma_{\tilde{\nu}_\tau}(\lambda^\prime_{333})$.

Finally, for $\lambda_{233}\ne 0$, $\tilde{\mu}$ and
$\tilde{\nu}_{\mu}$ may also be produced on shell in the decays
${\tilde{\chi}^0_1}\to \mu \bar{\tau}\nu_\tau$ and
${\tilde{\chi}^0_1}\to \nu_\mu \bar{\tau}\tau$. The width of the
$\tilde{\mu}$ slepton is then:
\begin{equation}
 \Gamma_{\tilde{\mu}_1}(\lambda_{233}) =   
 \Gamma(\tilde{\mu} \to \tau \nu_\tau)  =  
 \frac{1}{16 \pi}\vert\lambda^\prime_{333} \vert^2 \, 
  m_{\tilde{\mu}} \sin^2 \theta_\mu 
 \left(1 -\frac{m_\tau^2}{m_{\tilde{\mu}}^2} \right) 
  K^{1/2}\left(1,\frac{m_\tau}{m_{\tilde{\mu}}},0 \right),
\end{equation}
that for the sneutrino $\tilde{\nu}_{\mu}$ is 
\begin{equation}
 \Gamma_{\tilde{\nu}_\mu}(\lambda_{233}) =    
   \Gamma \left(\tilde{\nu}_\mu \to \overline{\tau_L} \,\tau_R \right) + 
   \Gamma \left(\tilde{\nu}_\mu \to \overline{\tau_R} \,\tau_L \right) = 
2\,\Gamma \left(\tilde{\nu}_\mu \to \overline{\tau_L} \,\tau_R \right) ,  
\end{equation}
with
\begin{equation}
\Gamma (\tilde{\nu}_\mu \to \overline{\tau_L} \,\tau_R) =
 \frac{1}{16 \pi}\vert\lambda^\prime_{333} \vert^2 \, 
  m_{\tilde{\nu}_\tau} 
 \left(1 -2 \frac{m_\tau^2}{m_{\tilde{\nu}_\tau}^2} \right) 
  K^{1/2}\left(1,\frac{m_\tau}{m_{\tilde{\nu}_\tau}},
                     \frac{m_\tau}{m_{\tilde{\nu}_\tau}} \right).
\end{equation}

\section{Spin sum of matrix elements squared} 
\label{squaredoperators} 

We list here the products of all possible $Q$ terms relevant for the
decays $\tilde{\chi}^0_1\to b \bar{b}\nu$ and $\tilde{\chi}^0_1\to t
\bar{b}\tau$ summed over all spin configurations.  These products are
evaluated under the assumption that the particle with momentum $p_1$
has a mass $m_1$, such that the ratio
\begin{equation}
 r = \frac{m_1^2}{m_{\tilde{\chi}^0_1}^2}
\end{equation}
is nonnegligible, as in the decay $\tilde{\chi}^0_1\to t \bar{b}
\tau$. When dealing with decays into a massless final state, as
$\tilde{\chi}^0_1\to b \bar{b}\nu$, it is sufficient to take the limit
$r \to 0$ in the following expressions.  For simplicity, the new
symbol
\begin{equation}
 \beta_{s,s^{\prime}}^{t,t^{\prime}} = \sum Q_{S,s}^{t}
 Q_{S,s}^{t^{\prime}\dagger}\,.
\label{betadef} 
\end{equation}
is introduced, with $t,t^{\prime} = x,y,z$ and where $s,s^{\prime}$
run over $RR$, $RL$, $LR$, and $LL$.
\begin{eqnarray}
\beta_{RR,RR}^{x,x} & = & m_{\tilde{\chi}^0_1}^4\,(1+r-x)x \nonumber
\\
\beta_{RR,RR}^{y,y} & = & m_{\tilde{\chi}^0_1}^4\,(1-y)(y-r)\nonumber
\\
\beta_{RR,RR}^{z,z} & = & m_{\tilde{\chi}^0_1}^4\,(1-z)(z-r)\nonumber
\\
\beta_{RL,RL}^{x,x} & = & m_{\tilde{\chi}^0_1}^4\,(1+r-x)x \nonumber
\\
\beta_{RL,RL}^{y,y} & = & m_{\tilde{\chi}^0_1}^4\,(1-y)(y-r)\nonumber
\\
\beta_{RL,RL}^{z,z} & = & m_{\tilde{\chi}^0_1}^4\,(1-z)(z-r) \,,
\label{spinsumoper}
\end{eqnarray} 
where $x$, $y$, and $z$ are defined in eq.~(\ref{xyzdefs}).  The same
results are obtained when the chirality indices in the above terms are
exchanged, i.e.\ 
\begin{eqnarray} 
\beta_{LL,LL}^{t,t} & = & \beta_{RR,RR}^{t,t}
\nonumber\\
\beta_{LR,LR}^{t,t} & = & \beta_{RL,RL}^{t,t}\,.
\end{eqnarray} 
Mixed products such as:
\begin{eqnarray} 
\beta_{RR,RR}^{x,y} & = & \frac{1}{2} m_{\tilde{\chi}^0_1}^4
\left[(1+r-x)x+(1-y)(y-r)-(1-z)(z-r)\right] 
\nonumber\\ 
\beta_{RR,RR}^{x,z} & = & \frac{1}{2} m_{\tilde{\chi}^0_1}^4
\left[(1+r-x)x-(1-y)(y-r)+(1-z)(z-r)\right] 
\nonumber\\ 
\beta_{RR,RR}^{y,z} & = & \frac{1}{2}m_{\tilde{\chi}^0_1}^4
\left[(1+r-x)x-(1-y)(y-r)-(1-z)(z-r)\right] 
\nonumber\\ 
\beta_{RL,RL}^{x,y} & = & 0 
\nonumber \\
\beta_{RL,RL}^{x,z} & = & 0 
\nonumber \\
\beta_{RL,RL}^{y,z} & = & 0 \,,
\label{spinsumoper_mixt}
\end{eqnarray} 
are invariant when chirality indices are exchanged, $R \leftrightarrow
L$, as they are the products
\begin{eqnarray} 
\beta_{RR,RL}^{x,x} & = & 2 m_{\tilde{\chi}^0_1}^4 \,\sqrt{r}\,x
\nonumber\\ 
\beta_{RR,RL}^{y,y} & = & 0 
\nonumber \\
\beta_{RR,RL}^{z,z} & = & 0 \,.
\label{spinsumoper_mixs}
\end{eqnarray} 
In addition, those in eq.~(\ref{spinsumoper_mixt}) are symmetric under
the exchange $t\leftrightarrow t^{\prime}$.  Finally, it is:
\begin{eqnarray} 
\beta_{RR,RL}^{x,y} & = & 0 
\nonumber \\
\beta_{RR,RL}^{x,z} & = & 0 
\nonumber \\
\beta_{RR,RL}^{y,x} & = & m_{\tilde{\chi}^0_1}^4\,\sqrt{r}\,x
\nonumber\\ 
\beta_{RR,RL}^{y,z} & = & 0 
\nonumber \\
\beta_{RR,RL}^{z,x} & = & m_{\tilde{\chi}^0_1}^4 \,\sqrt{r}\,x
\nonumber\\ 
\beta_{RR,RL}^{z,y} & = & 0
\label{spinsumoper_mixst}
\end{eqnarray} 
where these products are invariant under the exchange $R
\leftrightarrow L$ and are symmetric under the simultaneous exchange
$t\leftrightarrow t^{\prime}$ and $s\leftrightarrow s^{\prime}$,
\begin{equation}
 \beta_{s,s^{\prime}}^{t,t^{\prime}}  = 
 \beta_{s^{\prime},s}^{t^{\prime},t}\,.
\end{equation} 
 
Mixed products of terms with $s$ in the subset of indices $RR$, $RL$,
and $s^{\prime}$ in $LL$, $LR$, or vice versa, vanish identically,
except for the two terms
\begin{equation}
\beta_{LR,RL}^{y,z} = \beta_{RL,LR}^{y,z} = m_{\tilde{\chi}^0_1}^4
\,\sqrt{r}\,x \,.
\label{spinsumoper_mixnew}
\end{equation}


\end{document}